\documentclass[fleqn,10pt]{wlscirep}
\usepackage{amssymb,amsfonts,amsmath}

\newcommand{\ud}[1]{{#1^{\dagger}}}
\newcommand{\bra}[1]{\left\langle #1\right|}
\newcommand{\ket}[1]{\left| #1\right\rangle}

\newcommand{\mean}[1]{\langle #1\rangle}

\title{The colored Hanbury Brown--Twiss effect}

\author[1,2]{Blanca Silva}
\author[2]{C.~S\'anchez~Mu\~{n}oz}
\author[1]{D.~Ballarini}
\author[3]{A.~Gonz\'alez~Tudela}
\author[1]{M.~de~Giorgi}
\author[1]{G.~Gigli}
\author[4]{K.~W.~West}
\author[4]{L.~Pfeiffer}
\author[2]{E. del Valle}
\author[1]{D. Sanvitto}
\author[2,5,*]{F.~P.~Laussy}

\affil[1]{CNR NANOTEC--Institute of Nanotechnology, Via Monteroni, 73100 Lecce, Italy}
\affil[2]{Departamento de F\'isica Te\'orica de la Materia Condensada and Condensed Matter Physics Center (IFIMAC)}
\affil[3]{Max--Planck Institut f\"ur Quantenoptik, 85748 Garching, Germany}
\affil[4]{Department of Electrical Engineering, Princeton University, Princeton, New Jersey 08544, USA}
\affil[5]{Russian Quantum Center, Novaya 100, 143025 Skolkovo, Moscow Region, Russia}

\affil[*]{fabrice.laussy@gmail.com}

\begin{abstract}
  The Hanbury Brown--Twiss effect is one of the celebrated
  phenomenologies of modern physics that accommodates equally well
  classical (interferences of waves) and quantum (correlations between
  indistinguishable particles) interpretations. The effect was
  discovered in the late thirties with a basic observation of Hanbury
  Brown that radio-pulses from two distinct antennas generate signals
  on the oscilloscope that wiggle similarly to the naked eye.  When
  Hanbury Brown and his mathematician colleague Twiss took the obvious
  step to propose bringing the effect in the optical range, they met
  with considerable opposition as single-photon interferences were
  deemed impossible. The Hanbury Brown--Twiss effect is nowadays
  universally accepted and, being so fundamental, embodies many
  subtleties of our understanding of the wave/particle dual nature of
  light. Thanks to a novel experimental technique, we report here a
  generalized version of the Hanbury Brown--Twiss effect to include
  the frequency of the detected light, or, from the particle point of
  view, the energy of the detected photons. In addition to the known
  tendencies of indistinguishable photons to arrive together on the
  detector, we find that photons of different colors present the
  opposite characteristic of avoiding each others. We postulate that
  fermions can be similarly brought to exhibit positive (boson-like)
  correlations by frequency filtering.
\end{abstract}
\begin{document}

\flushbottom
\maketitle
%
%
\thispagestyle{empty}



\section*{Introduction}

The science of photon correlations---quantum optics---started with the
theory that Glauber developed to account for the conclusive
observation by Hanbury Brown and Twiss~\cite{hanburybrown56a} that
photons from thermal light detected at the single particle level do
indeed exhibit bunching in their arrival time, in the same way as
radio-waves correlated in intensities~\cite{hanburybrown52a}. The word
``coherent'' then changed from the meaning as used by
HBT~\cite{hanburybrown56b} (to mean monochromatic) to that of
Glauber~\cite{glauber63c} (to mean of uncorrelated photons). The fact
that initially unrelated photons, emitted maybe from different stars
in different galaxies, would exhibit a bunching effect, that is, a
tendency of arriving together on the detector, provoked much outrage
and incredulity in many of the prominent physicists of the
time~\cite{hanburybrown_book91a}, despite having an immediate
classical interpretation in terms of constructive
interferences~\cite{baym98a}.  This phenomenon was quickly understood
by Purcell~\cite{purcell56a} as, not only compatible with the particle
point of view, but also required by it, being associated to the
positive pair-correlation between bosons caused by their
indistinguishability. A complete formalization of the underlying
principle has been Nobel-prize winning~\cite{glauber06a}, culminating
with a now central quantity in quantum optics, the ``Glauber's
second-order coherence function~$g^{(2)}$'' defined as:
\begin{equation}
  \label{eq:lunmay19112320CEST2014}
  g^{(2)}(t,\tau)=
  \frac{\langle \hat E^-(t)\hat E^-(t+\tau) \hat E^+(t+\tau) \hat E^+(t) \rangle}{\langle \hat E^-(t) \hat E^+(t) \rangle  \langle \hat E^-(t+\tau) \hat E^+(t+\tau) \rangle}
\end{equation}
with~$\hat E^{+(-)}(t)$ the negative (positive) frequency part of the
Heisenberg electric field operator at time $t$ and $\tau$ the time
delay between detections (we omit position dependence for simplicity).
This quantity describes the statistical distribution between photons
in their stream of temporal detection.  Other properties of the
photons can be included, e.g., their position~\cite{hanburybrown56a}
or polarisation~\cite{stevenson06a}, with applications spanning from
atomic interferometry~\cite{jeltes07a} to entangled photon pair
generation~\cite{ekert91a}.

\section{Correlations when retaining the color of the photons}

Of all the possible additional variables that one can include or
retain when correlating the photons, one is so intertwined with the
temporal information as to define a special case of its own: it is the
energy of the photon (or, equivalently in the wave picture, its
frequency).  This is a characterisation of a different type than
position or polarisation, since time and frequency are conjugate
variables.  Frequency--resolved correlations are furthermore
observables that cannot be associated to a given quantum state, as
they also bring information on the dynamics of emission. This results
in a wider and unifying perspective of photon correlations. The formal
theory of time and frequency resolved correlations, established in the
80s~\cite{cohentannoudji79a, dalibard83a, knoll86a, nienhuis93a},
upgrades Eq.~(\ref{eq:lunmay19112320CEST2014}) to the two-photon
frequency correlations:
\begin{figure*}[t]
  \centering
 \includegraphics[width=0.9\textwidth]{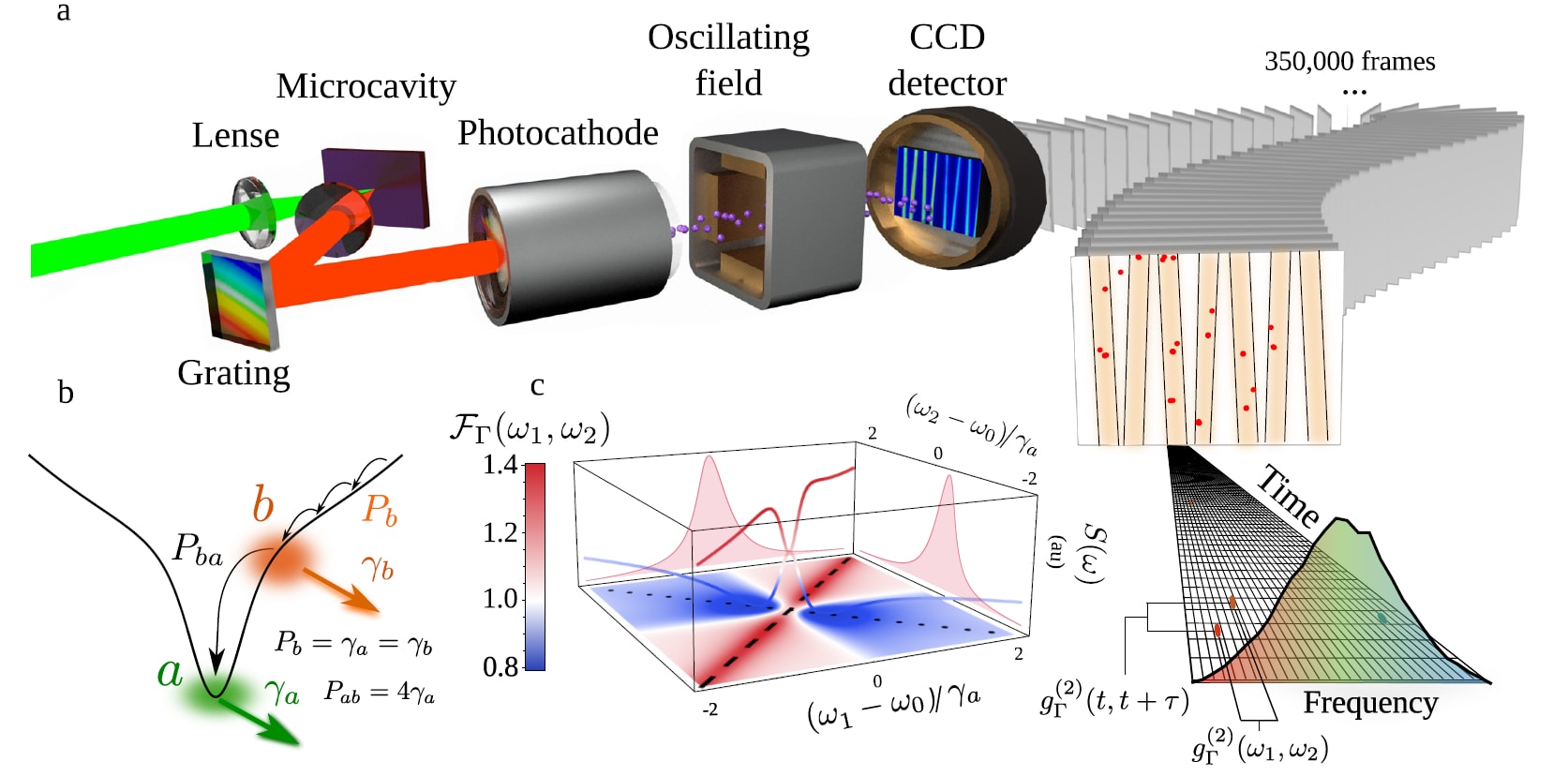}
  \caption{\textbf{Principle and Setup of time- and frequency-resolved
      photon correlations.} (a) Sketch of the experiment: the
    reflected light from a microcavity is dispersed onto a
    streak camera detecting at the single-photon level and stored in
    individual frames, whose post-processing allows to build
    photon-correlation landscapes (b) Sketch of the theory: a laser
    excites non-resonantly the lower polariton dispersion, creating a
    reservoir of hot excitons $b$ that condense into the ground state
    $a$ at the minimum of the branch. (c) Boson form factor
    $\mathcal{F}(\omega_1,\omega_2)$, i.e., time-integrated 2PS for
    the spontaneous emission of a coherent state with~$g^{(2)}=1$,
    providing the backbone for the experiment. The diagonal and
    antidiagonal exhibit bunching and antibunching, respectively.  \label{fig:1}}

\end{figure*}
\begin{equation}
  \label{eq:marjul29170631CEST2014}
  g_{\Gamma}^{(2)}(\omega_1,t_1;\omega_2,t_2)=\frac{\mean{:\mathcal{T}\big[\prod_{i=1}^2 \hat  E_{\omega_i,\Gamma}(t_i)\hat E^+_{\omega_i,\Gamma}(t_i)\big]:}}{\prod_{i=1}^2\mean{\hat E_{\omega_i,\Gamma}(t_i)\hat E^+_{\omega_i,\Gamma}(t_i)}},
\end{equation}
where
\begin{equation}
\hat E_{\omega_i,\Gamma}(t_i)= \frac{\Gamma}{2}\int_{0}^{\infty}
e^{-i\omega_i t}e^{-\Gamma t/2}\hat E(t_i-t)\,dt
\label{eq:frequency-filtered-field}
\end{equation}
is the electric field after passing through a filter with frequency
component $\omega_i$ and width $\Gamma$ at time $t_i$, and
$\mathcal{T}$, (resp.~$:$) refers to time (resp.~normal) ordering.
Equation~(\ref{eq:marjul29170631CEST2014}) provides the tendency of a
correlated detection of one photon of frequency~$\omega_1$ at
time~$t_1$ with another photon of frequency~$\omega_2$ at
time~$t_2$. We consider here Lorentzian filters but this discussion
applies to other types, such as square filters~\cite{kamide15a}.
Frequency-resolved photon correlations are an increasingly popular
experimental quantity, with already many measurements performed,
although for fixed sets of frequencies, merely by inserting filters in
the paths of a standard Hanbury Brown--Twiss setup~\cite{akopian06a,
  hennessy07a, kaniber08a, sallen10a, ulhaq12a, deutsch12a}. This
measurement reveals its conceptual importance, however, when spanning
over all possible combinations of energies, giving rise to a so-called
``two-photon correlation spectrum''
(2PS)~\cite{gonzaleztudela13a,peiris15a}.  Considering the most common
case of coincidences---$\tau=0$ in
Eq.~(\ref{eq:lunmay19112320CEST2014}) and $t_1=t_2$ in
Eq.~(\ref{eq:marjul29170631CEST2014})---one elevates in this way a
single number, ${g^{(2)}_0\equiv g^{(2)}(t=0,\tau=0)}$, to a full
landscape $g^{(2)}_\Gamma(\omega_1,\omega_2;\tau=0)$ of
correlations. The quantity defined by such a landscape (the 2PS)
acquires a fundamental meaning by revealing certain physical
features~\cite{gonzaleztudela13a,peiris15a}, in the same way that the
normal spectrum is meaningful because its an observable that spans
over a frequency range.

\section*{Results}

In this work, we report the complete HBT effect extended to the full
frequency-frequency map. We find that---in addition to the original
observation of positive correlations for identical photons reported by
the fathers of the effect and now routinely reproduced in a multitude
of quantum-optical platforms worldwide---photons also manifest
anticorrelations when they have different energies. Our results are a
direct extension of the original HBT effect, that is the particular
case of the diagonal on our 2PS. At such, it bears similar attributes
as well as counter-intuitive consequences. Namely, two photons
detected from two different sources manifest negative correlations if
detected in different frequency windows (namely, on opposite sides of
their mean energy) as compared to the unfiltered detection. If the
sources are coherent, so that the unfiltered detection presents no
correlation, the frequency-filtered photons exhibit anti-correlations:
the detection of photons of a given color makes it less likely to
detect photons of the other color. This behaviour is rooted
in the bosonic fabric through what we will introduce as the ``boson
form factor''. Like the original HBT effect, our findings can be
interpreted both from a quantum or a classical point of view, and
being due to boson statistics, represent a fundamental backbone of
every experiment involving frequency-resolved correlations. This
result is therefore of deep relevance for any measurement of this
kind, becoming of great interest in scenarios like the study of
fluctuations~\cite{sallen10a} or the harvesting and use of quantum
correlations that only emerge in the frequency-frequency
domain~\cite{ulhaq12a,deutsch12a,gonzaleztudela13a,
  gonzaleztudela15a}.

We have measured such anticorrelations between single photons emitted
from a macroscopic out-of-equilibrium ensemble of exciton-polaritons
pumped around the threshold of condensation. Polaritons are
strongly-coupled light-matter bosonic particles in a semiconductor
microcavity~\cite{kavokin_book11a}. Such a source is more convenient
than a laser because it is, for our purposes, essentially a laser
with a broad linewidth, thereby allowing the spectral filtering.  
Besides, they have enjoyed thorough studies of their coherence
properties, including at the quantum optical level~\cite{deng02a,assmann09a,adiyatullin15a}. The
experiment is based on a streak camera setup that detects individual
photons from the spontaneous emission of an ensemble of polaritons
maintained in a non-equilibrium steady state under cw excitation. This
is the first time that such a technique is used in the continuous
pumping regime. The setup is sketched in Fig.~\ref{fig:1}a: light
coming from the steady state of polaritons is dispersed by a
spectrometer and is directed into the streak camera that is able to
detect single photon events, as has already been demonstrated with
standard photon correlations in time domain only, under pulsed
excitation~\cite{wiersig09a}. The sweeping in time and dispersion in
energy allow the simultaneous recording of both the time and frequency
of each detected photon in successive frames that are post-processed
to calculate intensity correlations. Each frame includes several
sweeps: in Fig.~1a, 8 sweeps per frame are shown as vertical orange
stripes, with red dots indicating single photon events. Within each
sweep, a time of 1536 ps is spanned in the vertical direction (3.2 ps
per pixel), while the photon energy, obtained by coupling a
spectrometer to the streak camera, is measured as horizontal pixel
positions within each sweep (each sweep covers a total energy range of
456.7 $\mu\mathrm{eV}$, with 10.6 $\mu\mathrm{eV}$ per pixel). With
the time- and energy-range used in this experiment, the overall
temporal and energy resolution of the setup are of 10 ps and 70
$\mu\mathrm{eV}$, respectively.  Correlation landscapes are obtained
from coincidences between these clicks, with an average of
$\approx1.69$ clicks per sweep in a total of 350\,000 frames. All the
analysis is done with the raw data only: there is no
normalisation and the correlations go to~1 at long time
self-consistently.
\begin{figure}[t!]
  \centering
  \includegraphics[width=0.45\textwidth]{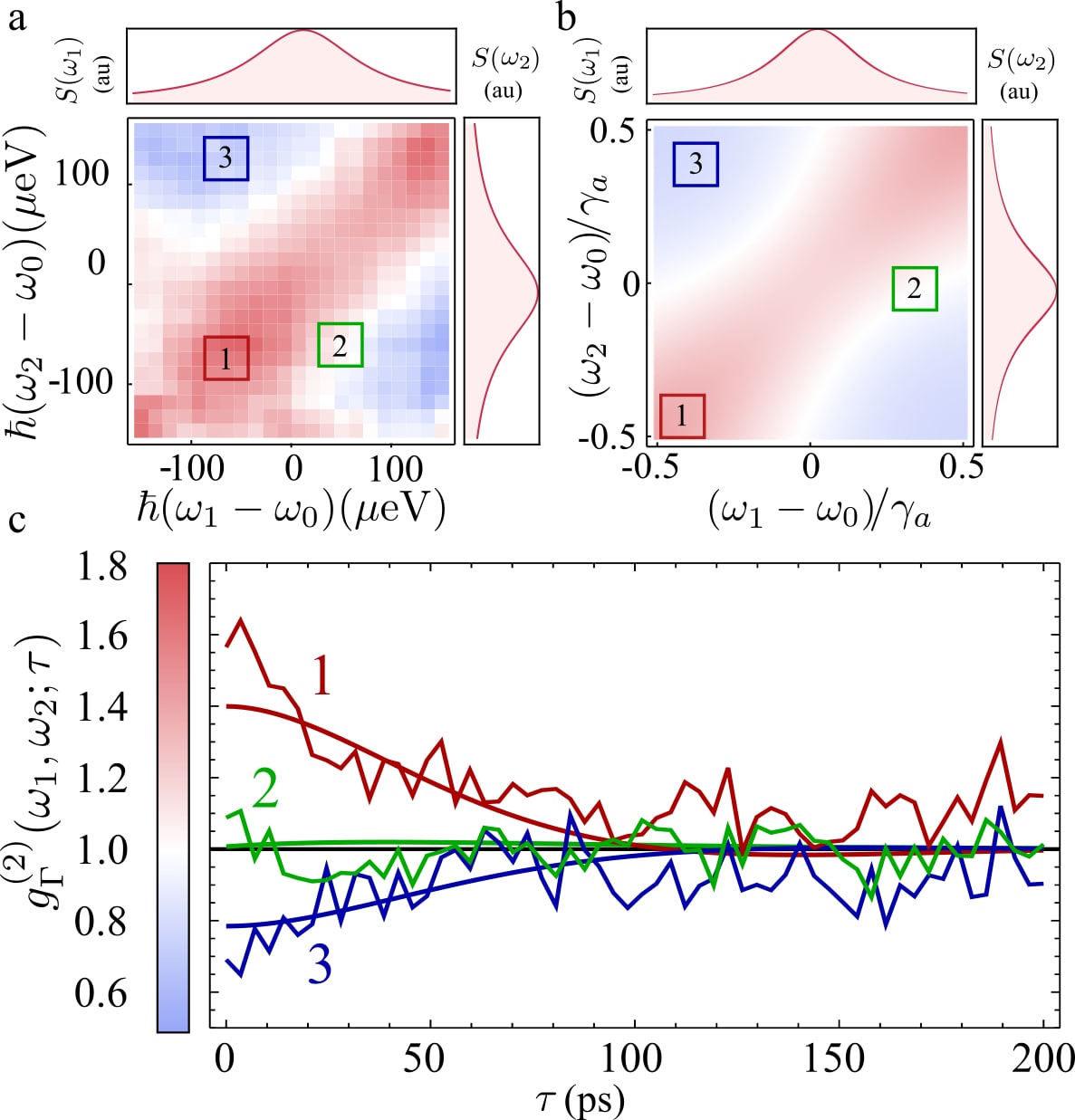}
  \caption{\textbf{Two-photon correlations spectra.} (a) Experimental
    observation of $g^{(2)}_\Gamma(\omega_1,\omega_2;0)$ for the
    spontaneous emission from a steady-state of polaritons. (b)
    Theoretical calculation of $g^{(2)}_\Gamma(\omega_1,\omega_2;0)$
    from the model of condensation of polaritons sketched in
    Fig.~\ref{fig:1}b., showing a remarkable agreement. (c)
    Time-resolved correlation for the three regions marked in the
    colour map: (i) on the diagonal ($\omega_1 = \omega_2$) exhibiting
    bunching, (ii) in the region of transition with no correlation,
    (iii) correlating opposing elbows, exhibiting antibunching.   \label{fig:2}}
\end{figure}
A scheme of the emitter is shown in Fig.~\ref{fig:1}b: polaritons
relax into the ground state from a reservoir of high energy polaritons
injected by a cw off-resonant laser. The constant losses through the
cavity mirror allow to study the steady-state correlations.  Both the
principle of the measurement and the technique are general and should
allow, with optimisation, to shed new light in already well known
systems in quantum optics, starting with the interesting non-classical
properties displayed by quantum sources. The outstanding time
resolution, which can be lower than 1 ps in a time window of 100 ps
per sweep, makes this technique also of interest to study other
phenomena such as spectral diffusion on fluctuating
systems~\cite{sallen10a}, promising to improve temporal precision by
two orders of magnitude.

Figure~\ref{fig:2}a shows the experimental 2PS for the polariton state
at $\tau=0$ together with the theoretical prediction, that is shown in
Fig.~\ref{fig:2}b and was computed from the steady state emission of
the model of condensation sketched in Fig.~\ref{fig:1}b using a master
equation and the recently developed sensors method~\cite{delvalle12a}
(see Section~IV of the Supplemental Material). Figure~\ref{fig:2}c
depicts the temporal correlations for three points of the
$(\omega_1,\omega_2)$--space, both for the experiment and the
theoretical model, demonstrating an outstanding time precision in the
scale of picoseconds. A clear evolution of the correlations from
bunching ($g^{(2)}_\Gamma(\omega,\omega;0)\approx1.5$ in region~1) to
antibunching ($g^{(2)}_\Gamma(-\omega,\omega;0)\approx0.7$ in
region~3) is observed.  In general, an excellent agreement with the
theory is obtained, especially for the salient features which are
diagonal bunching and antidiagonal antibunching.

\section*{Discussion}

\begin{figure*}[t!]
  \centering
 \includegraphics[width=1\textwidth]{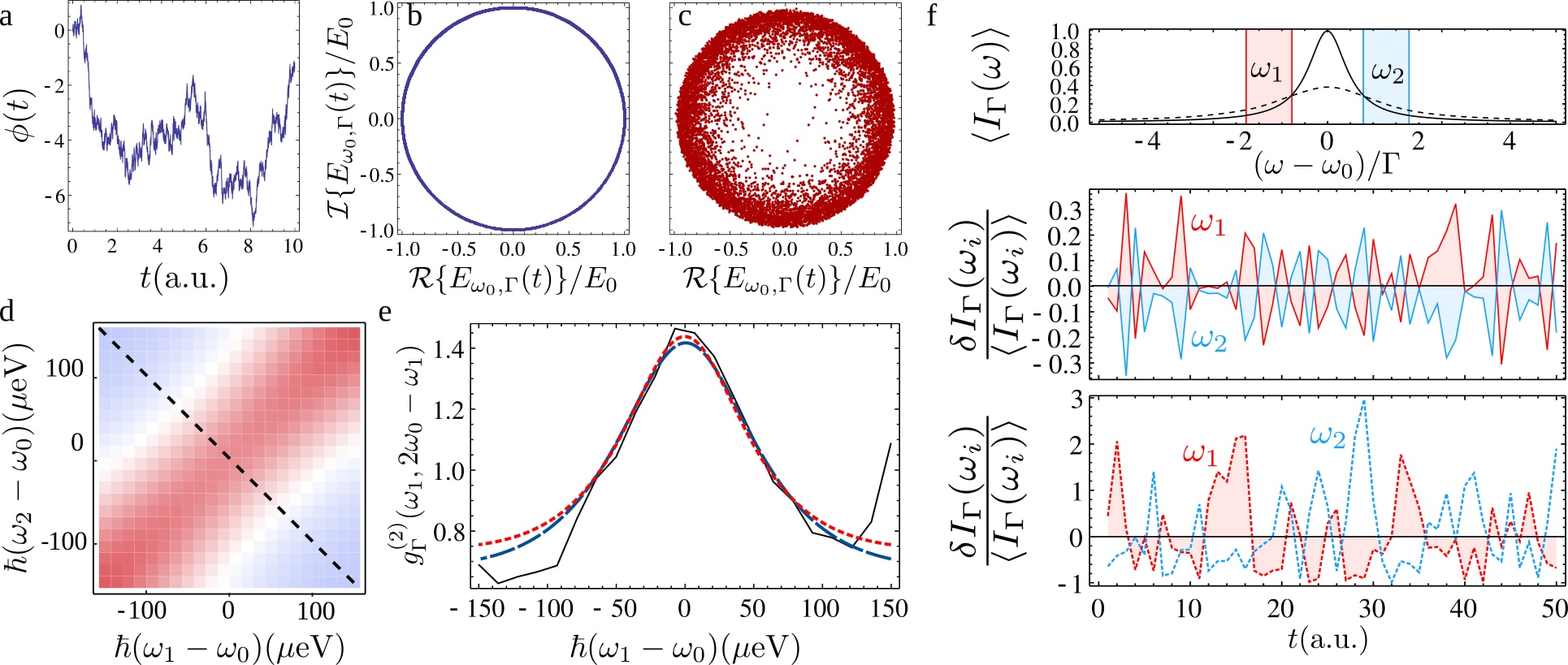}
 \caption{(a) The random-walk evolution of diffusing phase of a field
   $E(t) = E_0 \exp{i[\omega_0 t + \phi(t)]}$, with $\langle
   e^{i(\phi(t+\tau)-\phi(t))} \rangle = e^{- \gamma \tau}$ (b) $E(t)$
   in phase space over different times. (c) Phase fluctuations are
   converted into intensity fluctuations after frequency filtering
   $E(t)$ (d) Fitting of the experimental 2PS by
   equation~\eqref{eq:formula2ps}, with fitting parameters $\gamma
   \approx 193 \,\mu\mathrm{eV}$, $\Gamma \approx 134 \,
   \mu\mathrm{eV}$. The colorscale is that of Fig.~\ref{fig:2}. (e)
   2PS along the dashed line in (d) for the experiment (straight,
   black), the fitting for the phase diffusing field (long dashed,
   blue) and the fitting of the form factor
   $\mathcal{F}_{\Gamma,\gamma,\gamma_\phi}$ (short-dashed,
   red). Despite not being an exact theoretical description for this
   experiment, the form factor agrees very well with the data for the
   parameters $\gamma \approx 99\,\mu\mathrm{eV}$, $\gamma_\phi
   \approx 440 \,\mu\mathrm{eV}$, $\Gamma \approx 17
   \,\mu\mathrm{eV}$. (f) Fluctuations in the intensity of the
   filtered field $I_\Gamma(\omega_i) = \langle I_\Gamma(\omega_i)
   \rangle + \delta I_\Gamma(\omega_i) $ for the two frequencies shown
   at the top panel and two values of $\gamma$, $\gamma \approx
   8\times 10^{-3} \Gamma$ (solid lines, middle panel), and $\gamma
   \approx 0.8\, \Gamma$ (dashed lines, bottom panel). The
   corresponding values of $g^{(2)}(\omega_1,\omega_2)$ are 0.97 and
   0.65 resp. In the middle-panel case, where~$\Gamma\gg\gamma$,
   the anticorrelations in the noise become exact.   \label{fig:3}}
\end{figure*}

To the best of our knowledge, these features portray the first
evidence of a HBT effect generalized to the full frequency-frequency
domain. We now discuss these results in detail. Bunching in the
diagonal line (corresponding to filters of equal frequency) is the
well known feature of spectral filtering from a single
peak~\cite{centenoneelen93a
}. From a classical point of view, this can be understood with the
particular case of a quasi-monochromatic field $E(t)$ that has a
finite bandwith given by a phase diffusion process:
\begin{equation}
E(t) = E_0 e^{i[\omega_0 t + \phi(t)]},
\label{eq:phase-diphusing-field}
\end{equation}
where $\phi(t)$ is a stochastic function that evolves, for instance,
according to a random walk (see Fig.~\ref{fig:3}a). Such an errand
phase allows for the line broadening. As is clear from
Eq.~\eqref{eq:frequency-filtered-field}, the frequency-filtered field
is obtain by summing the field to itself at different times. If phase
diffusion is present, this corresponds to the superposition of fields
with random phases, which is analogous to the description of a thermal
field. Consequently, such a superposition of fields of equal frequency
but different phase produces interferences that wildly oscillate in a
chaotic intensity profile, resulting in:
\begin{equation}
\frac{\langle {I_\omega}^2 \rangle}{\langle I_\omega \rangle ^2} >1.
\label{eq:bunching}
\end{equation}
This is well known textbook material~\cite{loudon_book00a}.
Interpreted in terms of photons, the underlying particles thus tend to
``clump'' together, and increase the spacing between their arrival
time, which gives rise to the bunching effect.

We have just seen how phase noise is thus converted into intensity
noise by frequency-filtering (see Fig.~\ref{fig:3}b-c). In a related
but subtler way---which is the novel feature that we report--such
correlations can be negative when they involve different frequencies.
This remains true at the single particle level, as is demonstrated by
our experiment, with anticorrelations between photons of different
colors. Since the effect is linked to the aforementioned conversion of
phase noise into amplitude noise by filtering, we can keep the
paradigmatic case of a quasi-monochromatic field, that has only phase
noise. On physical grounds, one expects that a field with a stabilized
Poynting vector (in which the uncertainty in the number of photons
detected in a certain time window is given by the shot noise) cannot
yield in average more photon counting events per unit time when
spectrally resolved than it does without being frequency-filtered.
Therefore, the detection of a clump of photons of some frequency in a
small time window---in which photons are detected as random events
prior filtering---must lower the probability of detecting photons at
other, different frequencies, in order for the total rate of detected
photons to be preserved. The anticorrelation we observe can therefore
be interpreted as a consequence of energy conservation acting together
with the HBT effect, that yields bunching of indistinguishable photons
of equal frequencies. The photons on the detector, even if unrelated
in the first place, cannot afford to remain so when
frequency-filtered.  

This argument is verified by explicit computation of
Eq.~\eqref{eq:lunmay19112320CEST2014} applied on the field
Eq.~\eqref{eq:phase-diphusing-field}, assuming random walk dynamics
for the phase such that ${\langle e^{i[\phi(t)-\phi(t-\tau)]}\rangle}
= e^{-\gamma |\tau|}$, ${\langle e^{2i[\phi(t)-\phi(t-\tau)]}\rangle =
  e^{-\gamma_2 |\tau|}=e^{-4\gamma |\tau|}}$. The analytical
expression for the frequency resolved correlation function at zero
delay can be found exactly (the details of the calculation are given
in Section~I of the Supplemental Material):

\begin{multline}
g^{(2)}_\Gamma(\Delta_1,\Delta_2) = \frac{\left[\Delta_1^2+(\gamma+\Gamma/2)^2 \right] \left[\Delta_2^2+(\gamma+\Gamma/2)^2 \right] }{4(\gamma+\Gamma/2)^2}\times{}\\\Re \left\{ \frac{2(\gamma+3\Gamma/2)}{(\gamma+i\Delta_2+\Gamma/2)(\Delta_1^2+(\gamma+3\Gamma/2)^2)} +\Gamma\left[f_{\Gamma}(\Delta_2,\Delta_{12}^-,\Delta_2) \right. \right.
 +f_{\Gamma}(\Delta_1,\Delta_{12}^-,-\Delta_2)
 +f_{\Gamma}(\Delta_1,\Delta_{12}^+-i\gamma_2,\Delta_1)+\\
 \left. \left. 
 f_{\Gamma}(\Delta_2,\Delta_{12}^+-i\gamma_2,\Delta_1) \right] +
  \left[\Delta_1 \leftrightarrow \Delta_2\right]
  \vphantom{\frac{2(\gamma+3\lambda)}{(\Delta_1^2+(\gamma+3\lambda)^2)}}
  \right\}
 \label{eq:formula2ps}
\end{multline}

where $\Delta_i \equiv \omega_i-\omega_0$, $\Delta_{12}^- \equiv \Delta_2-\Delta_1$, $\Delta_{12}^+ \equiv \Delta_2+\Delta_1$ and:
\begin{equation}
  f_{\Gamma}(\omega_1,\omega_2,\omega_3) = \frac{1}{(i\omega_1+\gamma+\Gamma/2)(i \omega_2 + \Gamma)(i\omega_3+\gamma+3\Gamma/2)}\,.
\end{equation}
This expression reflects the same structure of correlations and
anticorrelations observed in the experiment, as depicted in
Fig.~\ref{fig:3}d-e, where it is shown to fit very well the
experimental data. The main assumption behind this equation--that the
unfiltered field has negligible amplitude fluctuations--is closely met
in the experiment, in which the high coherence degree of the light
emitted by the polaritons around the condensation threshold allows to
unambiguously observe the anticorrelations. Just as the
autocorrelations of Hanbury Brown for radio-waves of same frequencies
(with no filtering), these anticorrelations of the filtered signal are
obvious even to the naked eye, as can be seen in Fig.~\ref{fig:3}f,
showing the intensity fluctuations of the simulated phase-diffusing
field after frequency filtering. Surprisingly, such anticorrelations
in the noise can even become exact, when the filter linewidth becomes
much larger than the natural linewidth of the field ($\Gamma \gg
\gamma$). This is proved in Section~I of the Supplemental
Material. In this case, although $g^{(2)}(\omega_1,\omega_2)$ gets
closer to one (converging to the ``unfiltered'' result), the smaller
fluctuations $\delta I_\Gamma(\omega_i)$ around the mean value
$I_\Gamma(\omega_i) = \langle I_\Gamma(\omega_i) \rangle + \delta
I_\Gamma(\omega_i) $ tend to become perfectly anticorrelated for frequencies in opposite sides of the spectrum, $\delta
I_\Gamma(\omega_0-\omega) \approx - \delta I_\Gamma(\omega_0+\omega)$, as can be
observed in the middle panel of Fig.~\ref{fig:3}f.

Describing this effect from the quantum/particle point of view poses
more difficulties, since the 2PS is a dynamical observable and one
cannot attribute a value of~$g^{(2)}_0$ with frequency-filtering to a
given quantum state without also including information about the
dynamics, unlike the case without frequency-filtering where the
knowledge of the diagonal elements of the density matrix is
sufficient. Such differences are discussed from a more technical point
of view in Section~II of the Supplemental Material. This makes the
formulation of a general statement a complicated task. We consider for
that purpose a simple situation in which an arbitrary quantum state
given by the density matrix $\rho(0)$ is left to decay from a source
to a continuum of modes under spontaneous emission with a rate
$\gamma_a$ and also with a pure dephasing rate $\gamma_\phi$, thus
eliminating every possible dynamics except the essential one that
brings photons from the source to the detector and some dephasing
mechanism. Therefore, the resulting master equation is given by
$\partial_t\rho =
\left[\frac{\gamma_a}{2}\mathcal{L}_a+\frac{\gamma_\phi}{2}\mathcal{L}_{a^\dagger
    a} \right]\rho$, where $\mathcal{L}_O$ denotes the usual Lindblad
term, $\mathcal{L}_O\rho=2O\rho O^\dagger-O^\dagger O \rho - \rho
O^\dagger O $. We have obtained the analytical expression of the
normalized correlation for the counts at different frequencies
integrated in time, which takes the form:
\begin{equation}
g^{(2)}_\Gamma(\omega_1,\omega_2) = g_0^{(2)} \mathcal{F}_{\Gamma,\gamma,\gamma_\phi}(\omega_1,\omega_2)
\label{eq:g2-formfactor}
\end{equation}
with $g_0^{(2)}$ the zero delay second-order correlation function of
the initial state and
$\mathcal{F}_{\Gamma,\gamma,\gamma_\phi}(\omega_1,\omega_2)$ a form
factor (see Fig.~\ref{fig:1}c, Fig.~\ref{fig:3}e and Section~III of
the Supplemental Material for the analytical expression), independent
of this state, that reproduces exactly the features observed in the
experiment and therefore captures the essence of this extension of the
HBT effect. The wide range of frequencies used in Figure 1c serves to
illustrate the non trivial shape of the anticorrelations along the
antidiagonal line $(\omega,-\omega)$, featuring a minimum
approximately at the point where the total filtered intensity is
maximum without a considerable overlapping of the filters. It follows
from the arguments above that some dephasing mechanism is an essential
ingredient for the manifestation of the phenomenon (as it is for the
standard HBT effect). This is directly implied in the classical
picture and consistently confirmed in the quantum calculation, since
when the dephasing rate $\gamma_\phi$ is equal to zero,
$\mathcal{F}_{\Gamma,\gamma,\gamma_\phi}(\omega_1,\omega_2)$ is equal
to one, and therefore featureless. This also explains why the coherent
part of the resonance fluorescence spectrum, not subjected to
dephasing, does not present this features while the incoherent part
does (see Fig.~7 of \cite{gonzaleztudela13a}). Consistently with the
classical analysis based on a stochastic field
\eqref{eq:phase-diphusing-field}, the quantum calculation shows that
when the unfiltered field has no intensity fluctuations $g_0^{(2)} =
1$, the filtered field displays anticorrelations. Both general
analysis (classical and quantum), together with the experiment and the
theoretical characterization of the steady state emission of our
specific system, complete our description of the effect.
\begin{figure}[b!]
  \centering
 \includegraphics[width=0.45\textwidth]{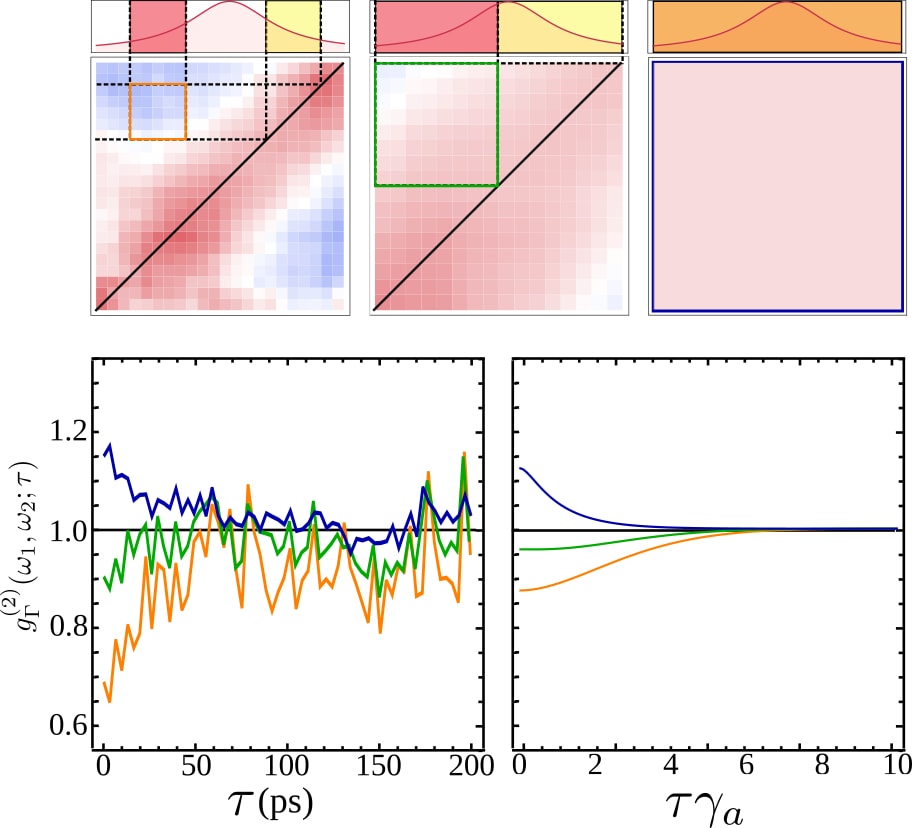}
 \caption{Two-photon correlation
   landscapes $g^{(2)}(\omega_1,\omega_2;0)$ as a function of the
   filter width, from a fraction of the peak, 74.1 $\mu eV$ (left), roughly half-peak width, 158.8 $\mu eV$
   (center), to full-peak filtering,
   corresponding to standard auto-correlations. The position of the
   two filters is shown explicitly on the spectral line as the red and
   yellow windows (orange when overlapping). Bottom left and right
   panels describe the experiment and the theory from the condensation
   model, respectively.}    
   \label{fig:4}
\end{figure}
Another fundamental feature of the theory is that correlations depend
on the frequency windows that select which photons are
correlated. Smaller windows lead to stronger correlations but, again,
at the price of a smaller signal. While it does not correspond exactly
to a change in the width of the filter, the effect is neatly
illustrated by changing the number of pixels of the streak camera that
we associate to a given frequency. In Fig.~\ref{fig:3}f, we show the
dependence of the 2PS on the size of the frequency windows for a point
that features antibunching. When the frequency window is very large,
$\Gamma\gg \gamma_a$, both the experimental and theoretical
$g^{(2)}_\Gamma(\omega_1,\omega_2;\tau)$ recover as expected the
results of standard photon correlations which has always been reported
to be larger than 1 for this kind of
systems~\cite{love08a,kasprzak08a,assmann11a}. As the size of the
frequency window decreases, the system shows a transition from
bunching to antibunching, demonstrating how the statistics of coloured
photons can be easily tuned externally.

These results, that generalize the Hanbury Brown--Twiss effect to
exhibit correlations of different types depending on the energies---
are of fundamental interest, but are also of technological importance.
Indeed, the main observable---the frequency-resolved correlation
function---is of increasing importance in quantum-optical
technologies.  Spectrally-resolved photon counting measurements can be
a useful tool in non-linear spectroscopy, able to measure ultrafast
dynamics~\cite{arXiv_dorfman16a}. Applications for measuring
fluctuations through detection of single photons in time and frequency
with picosecond resolution in spectral diffusion
problems~\cite{sallen10a} should also benefit from both our findings
and experimental setup. Our results and methods are also of importance
for the study of quantum systems with more complex dynamics beyond
merely spontaneous emission and
dephasing~\cite{delvalle12a,gonzaleztudela13a}. In cases of strongly
correlated emission, virtual
processes~\cite{sanchezmunoz14a,sanchezmunoz14b} result in 2PS with
strong geometric features, such as antidiagonals or circles of
correlations~\cite{gonzaleztudela13a,peiris15a}. Such rich landscapes
of photon correlations, inherited from the system's quantum dynamics,
are otherwise lost by disregarding the frequencies.  The results
presented here provide the backbone for more general schemes that, for
practical purposes such as distillation~\cite{delvalle13a} or Purcell
enhancement~\cite{sanchezmunoz14a,sanchezmunoz15a}, can be used to
power quantum technology. Like Purcell did in his pioneering
interpretation of the HBT effect~\cite{purcell56a}, we conjecture a
counterpart for fermion correlations, namely, with an orthogonal
profile: antibunching on the diagonal and bunching on the
antidiagonal; this question, that could be investigated for instance
in transport experiments with electrons~\cite{bocquillon14a}, is
however outside the scope of this text and its field of research, and
is left to experimental and theoretical colleagues in other
disciplines.

\section*{Conclusion}

We report the measurement of anticorrelations between individual
photons emitted from a ensemble of polaritons under continuous
pumping. We have demonstrated that this phenomenon is a fundamental
result that generalizes the Hanbury Brown--Twiss effect for color
correlations, and is therefore linked to the bosonic nature of
photons.  We have introduced a novel experimental technique that
allows to measure correlations in time and energy between individual
photons, demonstrating that both the concept and technique of color
correlations are sound and ripe to be deployed in a large range of
quantum optical systems, with prospects of accessing further classes
of quantum correlations~\cite{koch11a,rundquist14a}, optimising those
already known~\cite{delvalle13a,sanchezmunoz14b}, or analysing
problems such as spectral diffusion at new levels of
precision~\cite{sallen10a}.

\section*{Acknowledgements}

We thank G.~Guirales Arredondo and J.~C.~L\'opez Carre\~no for
discussions.  This work has been funded by the ERC Grant POLAFLOW
Project No. 308136., the IEF project SQUIRREL (623708) and by the
Spanish MINECO under contract FIS2015-64951-R (CLAQUE).  AGT
acknowledges support from the Alexander Von Humboldt Foundation,
C.S.M. from a FPI grant (MAT2011-22997, MAT2014- 53119-C2-1-R) and
F.P.L.~a RyC contract. The work at Princeton University was funded by
the Gordon and Betty Moore Foundation through the EPiQS initiative
Grant GBMF4420, and by the National Science Foundation MRSEC Grant
DMR-1420541.



\section*{Supplementary Material}

\subsection*{Materials}

The sample is a high quality factor GaAs/AlGaAs planar cavity
containing 12 GaAs quantum wells placed at three anti-node positions
of the electrical field. The front (back) mirror consists of 34 (40)
pairs of AlAs/Al0.2Ga0.8As layers. It is pumped at normal incidence
and non-resonantly with a single-mode laser at a wavelength of
752~$\mu\mathrm{m}$. Thanks to a double electronic synchronization, an
additional ``slow'' sweeping in time is performed also in the
horizontal direction, thus recording multiple sweeps per frame: in
this way the number of events detected in each frame is increased of
about one order of magnitude, allowing the detection of a large
statistics of events in a relatively short measurement time, going
beyond the limits imposed by the electronics speed of CCD devices.

\subsection*{I.~Calculation of $g^{(2)}_\Gamma(\omega_1,\omega_2,\tau)$ for a phase diffusing field}

In this section, we give the details of the calculation of the
frequency-resolved correlation function for a classical stochastic
field. Such a phase-diffusing field but with otherwise stabilized
intensity (describing a coherent field) reads:
\begin{equation}
E(t) = E_0 e^{-i[\omega_0 t + \phi(t)]}\,,
\end{equation}
where the phase is a random variable that is considered to undergo a
random walk evolution. The phase difference $\Delta \phi(\tau) =
\phi(t+\tau)-\phi(t)$ has the following properties:
\begin{eqnarray}
\langle \Delta \phi (\tau) \rangle &=& 0\,, \nonumber\\
\langle \Delta\phi(\tau)^2 \rangle &=& 2\gamma_1 |\tau|\,, \nonumber\\
\langle e^{i[\phi(t)-\phi(t-\tau)]} \rangle &=& e^{-\gamma_1 |\tau|}\,, \nonumber\\
\langle e^{2i[\phi(t)-\phi(t-\tau)]}\rangle &=& e^{-\gamma_2|\tau|}\,.
\label{eq:phase-averages}
\end{eqnarray}
In the case of a phase diffusing field, the fourth orther correlation
constant $\gamma_2$ is given by $\gamma_2 = 4\gamma_1$. However, we
keep $\gamma_2$ in the expressions to account for other possible
models of phase noise, like the phase-jump model, in which $\gamma_2 =
\gamma_1$~\cite{centenoneelen92a}. These properties allow us to
compute the frequency resolved second-order correlation function:
\begin{equation}
g^{(2)}_\Gamma (\omega_1,\omega_2,\tau) = \frac{\langle E_{\omega_2,\Gamma}^+(t+\tau)   E_{\omega_2,\Gamma}(t+\tau) E_{\omega_1,\Gamma}^+(t)   E_{\omega_1,\Gamma}(t)\rangle}{\langle E_{\omega_2,\Gamma}^+(t+\tau)   E_{\omega_2,\Gamma}(t+\tau) \rangle \langle E_{\omega_1,\Gamma}^+(t)   E_{\omega_1,\Gamma}(t) \rangle},
\label{eq:g2-freq}
\end{equation}
where $E_{\omega,\Gamma}(t)$ describes the field filtered at frequency
$\omega$ with a filter linewidth $\Gamma$:
\begin{equation}
E_{\omega,\Gamma}(t) = \frac{\Gamma}{2}\int_{0}^{\infty} e^{-i\omega t'}e^{-\Gamma\, t'/2}E(t-t')\,dt'.
\end{equation}
This calculation is done with the properties listed in
Eq.~\eqref{eq:phase-averages}. The numerator of
Eq.~\eqref{eq:g2-freq}, that we denote $G^{(2)}_\Gamma
(\omega_1,\omega_2,\tau)$, is given by the following quadruple
integral:
\begin{multline}
G^{(2)}_\Gamma (\omega_1,\omega_2,\tau) = \left(\frac{\Gamma}{2} \right)^4\int_0^\infty \prod_{i=1}^{4}dt_i \,e^{i \omega_1 (t_2-t_1)}e^{i\omega_2(t_4-t_3)} e^{-\Gamma(t_1+t_2+t_3+t_4)/2}\langle E(t-t_3+\tau) E^*(t-t_4+\tau)   E(t-t_1) E^*(t-t_2) \rangle \\
= E_0^4\left(\frac{\Gamma}{2} \right)^4\int_0^\infty\prod_{i=1}^{4}dt_i \,
e^{-i\Delta_1(t_2-t_1)-i\Delta_2(t_3-t_4)}
   e^{-\Gamma(t_1+t_2+t_3+t_4)/2}
   \langle e^{-i[\phi(t-t_1)-\phi(t-t_2)+\phi(t-t_3+\tau)-\phi(t-t_4+\tau)]} \rangle\,,
\label{eq:G2}
\end{multline}
where $\Delta_i \equiv \omega_0-\omega_i$.  Defining $t_1' \equiv
t-t_1$, $t_2' \equiv t-t_2$, $t_3' \equiv t-t_3+\tau$ and $t_4' \equiv
t-t_4+\tau$, the statistical average in the last line of \eqref{eq:G2}
takes the form $\langle
e^{i[\phi(t_1')-\phi(t_2')+\phi(t_3')-\phi(t_4')]} \rangle$. The
exponent can be written in term of phase differences
$\Delta\phi(\tau)$ in two possible ways, $\langle
e^{i[\Delta\phi(t_1'-t_2')+\Delta\phi(t_3'-t_4')]} \rangle$ or
$\langle e^{i[\Delta\phi(t_1'-t_4')+\Delta\phi(t_3'-t_2')]}
\rangle$. For a given set of values for $t_1'$, $t_2'$, $t_3'$ and
$t_4'$, the choice between both options must be made such that the two
$\Delta\phi$ are defined in non-overlapping time intervals, making
them statistically independent. This allows to factorize the
exponential and use Eq.~\eqref{eq:phase-averages} to evaluate the
statistical averages. Figure~\ref{fig:1} depicts the three possible
configurations that exist depending on the values of $t_i'$. Panel c
shows the particular case $t'_4,t_2'<t_1',t_3'$ that requires the
introduction of a third time interval to avoid overlapping; this is
the case that will invoke the last equation in
Eq.~\eqref{eq:phase-averages}, involving the fourth order correlation
constant $\gamma_2$. Since the integrand has to be written differently
depending on the values of the $t_i'$, one needs to split the integral
in the 24 possible domains. Half of these integrals are the complex
conjugate of the other half, yielding 12 independent terms, defined in
the domains:
\begin{eqnarray}
\mathrm{I:} &\quad& t_2' < t_1' < t_3' < t_4' \quad \stackrel{*}{\leftrightarrow} \quad t_1' < t_2' < t_4' < t_3'\nonumber\,,\\
\mathrm{II:} &\quad& t_2' < t_3' < t_1' < t_4' \quad \stackrel{*}{\leftrightarrow} \quad t_1' < t_4' < t_2' < t_3'\nonumber\,,\\
\mathrm{III:} &\quad& t_4' < t_3' < t_1' < t_2' \quad \stackrel{*}{\leftrightarrow} \quad t_3' < t_4' < t_2' < t_1'\nonumber\,,\\
\mathrm{IV:} &\quad& t_4' < t_1' < t_3' < t_2' \quad \stackrel{*}{\leftrightarrow} \quad t_3' < t_2' < t_4' < t_1'\nonumber\,,\\
\mathrm{V:} &\quad& t_2' < t_1' < t_4' < t_3' \quad \stackrel{*}{\leftrightarrow} \quad t_1' < t_2' < t_3' < t_4'\nonumber\,,\\
\mathrm{VI:} &\quad& t_2' < t_3' < t_4' < t_1' \quad \stackrel{*}{\leftrightarrow} \quad t_1' < t_4' < t_3' < t_2'\nonumber\,,\\
\mathrm{VII:} &\quad& t_4' < t_3' < t_2' < t_1' \quad \stackrel{*}{\leftrightarrow} \quad t_3' < t_4' < t_1' < t_2'\nonumber\,,\\
\mathrm{VIII:} &\quad& t_4' < t_1' < t_2' < t_3' \quad \stackrel{*}{\leftrightarrow} \quad t_3' < t_2' < t_1' < t_4'\nonumber\,,\\
\mathrm{IX:} &\quad& t_2' < t_4' < t_1' < t_3' \quad \stackrel{*}{\leftrightarrow} \quad t_1' < t_3' < t_2' < t_4'\nonumber\,,\\
\mathrm{X:} &\quad& t_2' < t_4' < t_3' < t_1' \quad \stackrel{*}{\leftrightarrow} \quad t_1' < t_3' < t_2' < t_4'\nonumber\,,\\
\mathrm{XI:} &\quad& t_4' < t_2' < t_1' < t_3' \quad \stackrel{*}{\leftrightarrow} \quad t_3' < t_1' < t_2' < t_4'\nonumber\,,\\
\mathrm{XII:} &\quad& t_4' < t_2' < t_3' < t_1' \quad \stackrel{*}{\leftrightarrow} \quad t_3' < t_1' < t_4' < t_2'\,.
\end{eqnarray}

By denoting the non-overlapping time differences as $\tau_i$ and making a change of variables, the corresponding integrals take the form:
\begin{eqnarray}
I_\mathrm{I} &=&C \int_0^\infty dt_2 \int_{t_2}^0 d\tau_1 \int_0^{t_2-\tau_1+\tau}  \kern-0.5cm  dt_3 \int_{t_3}^0 d\tau_2\, e^{i(\Delta_2 \tau_2-\Delta_1 \tau_1) }e^{-\Gamma(2t_2+2t_3-\tau_1 -\tau_2)/2 - \gamma(\tau_1+\tau_2)} \nonumber,\\
I_\mathrm{II} &=&C 
\int_0^\infty dt_2 
\int_{t_2}^0 d\tau_1 
\int_0^{t_2-\tau_1}  \kern-0.4cm  dt_1 
\int_{t_1+\tau}^0 d\tau_2\,
 e^{i\Delta_1(t_1-t_2)+i\Delta_2(\tau_2-\tau_1+t_2-t_1)  }e^{-\Gamma(2t_2-\tau_1+2t_1-\tau_2+2\tau)/2 - \gamma(\tau_1+\tau_2)}\nonumber,\\
I_\mathrm{III} &=&C 
\int_\tau^\infty dt_4 
\int_{t_4-\tau}^0 d\tau_1 
\int_0^{t_4-\tau_1-\tau}  \kern-0.5cm  dt_1 
\int_{t_1}^0 d\tau_2\,
 e^{i(\Delta_1 \tau_2 - \Delta_2 \tau_1 }e^{-\Gamma(2 t_4 + 2t_1 - \tau_1 - \tau_2)/2 - \gamma(\tau_1+\tau_2)}\nonumber,\\
I_\mathrm{IV} &=&C 
\int_\tau^\infty dt_4 
\int_{t_4-\tau}^0 d\tau_1 
\int_\tau^{t_4-\tau_1} \kern-0.5cm  dt_3 
\int_{t_3-\tau}^0 d\tau_2\,
 e^{i\Delta_2(t_3-t_4) + i\Delta_1(\tau_2-\tau_1 + t_4-t_3) }e^{-\Gamma(2t_4+2t_3-\tau_1 -\tau_2 - 2\tau)/2 - \gamma(\tau_1+\tau_2)} \nonumber,\\
I_\mathrm{V} &=&C 
\int_0^\infty dt_2 
\int_{t_2}^0 d\tau_1 
\int_0^{t_2-\tau_1+\tau} \kern-0.6cm  dt_4 
\int_{t_4}^0 d\tau_2\,
 e^{-i(\Delta_1 \tau_1 + \Delta_2 \tau_2) }
 e^{-\Gamma(2t_2+2t_4-\tau_1 -\tau_2)/2 - \gamma(\tau_1+\tau_2)} \nonumber,\\ 
I_\mathrm{VI} &=&C 
\int_0^\infty dt_2 
\int_{t_2}^0 d\tau_1 
\int_\tau^{t_2-\tau_1+\tau} \kern-0.6cm  dt_4 
\int_{t_4-\tau}^0 d\tau_2\,
 e^{-i\Delta_1(t_2-t_4+\tau+\tau_2) -i \Delta_2(t_4-t_2-\tau+\tau_1 ) }
 e^{-\Gamma(2t_2+2t_4-\tau_1 -\tau_2)/2 - \gamma(\tau_1+\tau_2)} \nonumber,\\ 
I_\mathrm{VII} &=&C 
\int_\tau^\infty dt_4 
\int_{t_4-\tau}^0 d\tau_1 
\int_0^{t_4-\tau_1-\tau}  \kern-0.5cm dt_2 
\int_{t_2}^0 d\tau_2\,
 e^{-i\Delta_1 \tau_2 - i\Delta_2 \tau_1}
 e^{-\Gamma(2t_4+2t_2-\tau_1 -\tau_2)/2 - \gamma(\tau_1+\tau_2)} \nonumber,\\ 
I_\mathrm{VIII} &=&C 
\int_\tau^\infty dt_4 
\int_{t_4-\tau}^0 d\tau_1 
\int_0^{t_4-\tau_1-\tau}  \kern-0.5cm dt_2 
\int_{t_2+\tau}^0 d\tau_2\,
 e^{-i\Delta_1(t_2-t_4+\tau+\tau_1)-i\Delta_2(t_4-t_2-\tau+\tau_2)}
 e^{-\Gamma(2t_4+2t_2-\tau_1 -\tau_2)/2 - \gamma(\tau_1+\tau_2)} \nonumber,\\ 
I_\mathrm{IX} &=&-C 
\int_0^\infty dt_2 
\int_{t_2}^0 d\tau_1 
\int_{t_2-\tau_1}^0  \kern-0.5cm d\tau_2 
\int_{t_2-\tau_1-\tau_2+\tau}^0  \kern-1cm d\tau_3\,
 e^{-i\Delta_1(\tau_1+\tau_2)-i\Delta_2(\tau_2+\tau_3)}
 e^{-\Gamma(4t_2-3\tau_1-2\tau_2-\tau_3+2\tau)/2 - \gamma(\tau_1+\tau_3)-\gamma_2 \tau_2} \nonumber,\\  
I_\mathrm{X} &=&-C 
\int_0^\infty dt_2 
\int_{t_2}^0 d\tau_1 
\int_{t_2-\tau_1}^0 d\tau_2 
\int_{t_2-\tau_1-\tau_2}^0  \kern-0.7cm d\tau_3\,
 e^{-i\Delta_1(\tau_1+\tau_2+\tau_3)-i\Delta_2 \tau_2}
 e^{-\Gamma(4t_2 + 2\tau - 3\tau_1 - 2\tau_2 - \tau_3)/2 - \gamma(\tau_1+\tau_3)-\gamma_2 \tau_2} \nonumber,\\   
I_\mathrm{XI} &=&-C 
\int_\tau^\infty dt_4 
\int_{t_4-\tau}^0 d\tau_1 
\int_{t_4-\tau-\tau_1}^0  \kern-0.5cm  d\tau_2 
\int_{t_4-\tau_1-\tau_2}^0  \kern-0.7cm d\tau_3\,
 e^{-i\Delta_1 \tau_2 -i\Delta_2(\tau_1+ \tau_2 + \tau_3)}
 e^{-\Gamma(4 t_4 - 2\tau - 3\tau_1 - 2\tau_2 - \tau_3)/2 - \gamma(\tau_1+\tau_3)-\gamma_2 \tau_2} \nonumber,\\   
I_\mathrm{XII} &=&-C 
\int_\tau^\infty dt_4 
\int_{t_4-\tau}^0 d\tau_1 
\int_{t_4-\tau-\tau_1}^0  \kern-0.5cm d\tau_2 
\int_{t_4-\tau_1-\tau_2-\tau}^0 \kern-1cm d\tau_3\,
 e^{-i\Delta_1 (\tau_2 + \tau_3) -i\Delta_2(\tau_1+ \tau_2)}
 e^{-\Gamma(4 t_4 - 2\tau -3\tau_1 -2\tau_2 - \tau_3)/2 - \gamma(\tau_1+\tau_3)-\gamma_2 \tau_2}\nonumber,\\
{} 
 \end{eqnarray}
with $C=E_0^4\left(\frac{\Gamma}{2} \right)^4$.
\begin{figure}[t]
  \centering
 \includegraphics[width=0.99\textwidth]{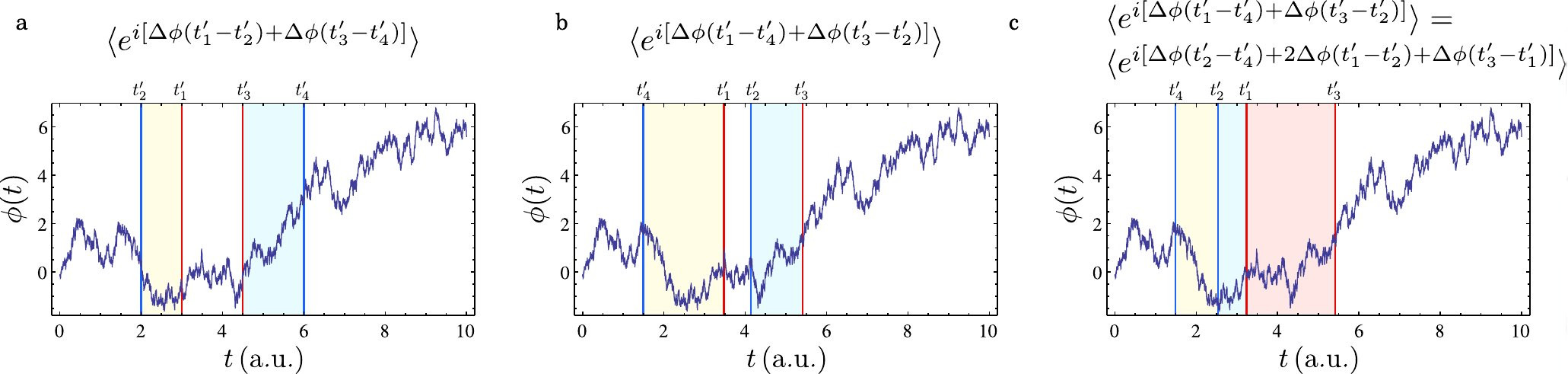}
 \caption{Examples of the possible integration domains for the
   fluctuating phase corresponding to three possible exponents that
   appear in the integrals: (a) Domain I, (b) Domain VIII and (c)
   Domain XI. In domains IX--XII, the exponent must be written as
   three phase differences to ensure they are statistically
   uncorrelated. }
  \label{fig:1}
\end{figure}
Finally, $G^{(2)}_\Gamma (\omega_1,\omega_2,\tau)$ is given by:
\begin{small}
\begin{eqnarray}
&&G^{(2)}_\Gamma (\omega_1,\omega_2,\tau)  = 2\mathrm{Re}\sum_{i=\mathrm{I}}^{\mathrm{XII}}I_i  = {}\nonumber\\
&&{}\nonumber\\
&& \frac{1}{16} \Gamma ^2 \mathrm{Re}  \left\{ \frac{2 i e^{\tau  \left(-\gamma -\frac{\Gamma }{2}-i
   \Delta _2\right)} \left(\gamma _2 \left(\frac{i \Gamma }{2}+2 \Delta _1-\Delta
   _2\right)+\left(\frac{\Gamma }{2}+i \Delta _1\right) \left(-i \Gamma +\Delta
   _1+\Delta _2\right)+\gamma  \left(-3 i \Gamma -i \gamma _2-\Delta _1+3 \Delta
   _2\right)\right) \Gamma ^2}{\left(\gamma +\frac{\Gamma }{2}+i \Delta _1\right)
   \left(\gamma -\frac{\Gamma }{2}+i \Delta _2\right) \left(\gamma +\frac{\Gamma
   }{2}+i \Delta _2\right) \left(\gamma +\frac{3 \Gamma }{2}+i \Delta _2\right)
   \left(-i \Gamma -\Delta _1+\Delta _2\right) \left(-i \Gamma -i \gamma _2+\Delta
   _1+\Delta _2\right)} \right. \nonumber \\  
   &+& \frac{2 e^{\tau  \left(-\gamma -\frac{\Gamma }{2}+i \Delta
   _2\right)} \left(\gamma -\frac{\Gamma }{2}-i \Delta _1\right) \Gamma
   ^2}{\left(\gamma +\frac{\Gamma }{2}+i \Delta _1\right) \left(\gamma +\frac{\Gamma
   }{2}-i \Delta _2\right) \left(\gamma +\frac{3 \Gamma }{2}-i \Delta _2\right)
   \left(i \gamma -\frac{i \Gamma }{2}+\Delta _2\right) \left(i \Gamma -\Delta
   _1+\Delta _2\right)}\nonumber \\
  &+& \frac{e^{-\Gamma  \tau } \left(-\Gamma ^3-\frac{1}{2} \left(4
   \gamma +i \left(3 \Delta _1+\Delta _2\right)\right) \Gamma ^2+\frac{1}{2}
   \left(\Delta _1-\Delta _2\right) \left(2 i \gamma -3 i \gamma _2+\Delta _1+3
   \Delta _2\right) \Gamma +\left(\Delta _1-\Delta _2\right) \left(\gamma +i \Delta
   _2\right) \left(-i \gamma _2+\Delta _1+\Delta _2\right)\right)}{\left(\gamma
   +\frac{\Gamma }{2}+i \Delta _1\right) \left(\gamma -\frac{\Gamma }{2}+i \Delta
   _2\right) \left(\gamma +\frac{\Gamma }{2}+i \Delta _2\right) \left(-i \Gamma
   -\Delta _1+\Delta _2\right) \left(-i \Gamma -i \gamma _2+\Delta _1+\Delta
   _2\right)}\nonumber \\
	&+&   
   e^{-\Gamma  \tau } \left[\frac{2 \left(\gamma +\frac{3 \Gamma
   }{2}\right)}{\left(\left(\gamma +\frac{3 \Gamma }{2}\right)^2+\Delta _1^2\right)
   \left(\gamma +\frac{\Gamma }{2}+i \Delta _2\right)}
   +\frac{i \Gamma }{\left(\gamma +\frac{3 \Gamma }{2}-i \Delta _1\right) \left(i \Gamma
   +\Delta _1-\Delta _2\right) \left(\gamma +\frac{\Gamma }{2}+i \Delta
   _2\right)}  \right.\nonumber \\
  &+&   \frac{\Gamma}{\left(\gamma +\frac{\Gamma }{2}+i \Delta _1\right)
   \left(\gamma +\frac{3 \Gamma }{2}+i \Delta _1\right) \left(\Gamma +\gamma _2+i
   \left(\Delta _1+\Delta _2\right)\right)}+\frac{\Gamma}{\left(\gamma +\frac{3 \Gamma
   }{2}+i \Delta _1\right) \left(\gamma +\frac{\Gamma }{2}+i \Delta _2\right)
   \left(\Gamma +\gamma _2+i \left(\Delta _1+\Delta
   _2\right)\right)}\nonumber \\
  &+&  \left.\frac{\Gamma}{\left(\gamma +\frac{\Gamma }{2}+i \Delta _1\right)
   \left(\gamma +\frac{3 \Gamma }{2}+i \Delta _1\right) \left(\Gamma +i \left(\Delta
   _1-\Delta _2\right)\right)}\right]+\frac{2}{\left(\gamma +\frac{\Gamma
   }{2}+i \Delta _1\right) \left(\gamma +\frac{\Gamma }{2}-i \Delta
   _2\right)}+\frac{2}{\left(\gamma +\frac{\Gamma }{2}+i \Delta _1\right)
   \left(\gamma +\frac{\Gamma }{2}+i \Delta _2\right)}\nonumber \\
   &+&\left. \frac{e^{-\Gamma  \tau }
   \left(\Delta _2-\Delta _1\right)}{\left(\gamma +\frac{\Gamma }{2}+i \Delta
   _1\right) \left(-\gamma +\frac{\Gamma }{2}+i \Delta _2\right) \left(i \Gamma
   -\Delta _1+\Delta _2\right)}\right\}\,.
   \label{eq:G2-long}
\end{eqnarray}
\end{small}
On the other hand, the denominator in Eq.~\eqref{eq:g2-freq} consists
of the product of the mean intensities of the two filtered
fields. This mean intensity is readily given by:
\begin{eqnarray}
\langle E^+_{\omega_i,\Gamma}(t)E_{\omega_i,\Gamma}(t) \rangle = \langle I_{\omega_i,\Gamma}(t) \rangle = E_0^2 \frac{\Gamma^2}{4}\left( \int_0^\infty dt_1 \int_0^{t_1} dt_2 \,e^{i\Delta_i (t_1-t_2) - \Gamma(t_1+t_2)/2 - \gamma(t_1-t_2)}  
+ \right.\nonumber \\
\left. \int_0^\infty dt_2 \int_0^{t_2} dt_1 \,e^{i\Delta_i (t_1-t_2) - \Gamma(t_1+t_2)/2 - \gamma(t_2-t_1)} \right) \nonumber\\
= E_0^2 \frac{\Gamma^2}{2} \mathrm{Re} \int_0^\infty dt_1 \int_0^{t_1} dt_2 \,e^{(i\Delta_i-\gamma)(t_1-t_2)-\Gamma(t_1+t_2)/2} = E_0^2\frac{\Gamma}{2}\frac{\gamma+\Gamma/2}{\Delta_i^2 + (\gamma + \Gamma/2)^2}\,.
\label{eq:intensity}
\end{eqnarray}
Normalizing the expression of $G^{(2)}_\Gamma
(\omega_1,\omega_2,\tau)$ given in Eq.~\eqref{eq:G2-long} by the
intensities in Eq.~\eqref{eq:intensity}, we obtain the final
expression for $g^{(2)}_\Gamma (\omega_1,\omega_2,\tau)$. At $\tau =
0$, this expression takes the more compact form presented in the main
text.

An interesting limit, reported in Fig.~3f of the main text, occurs
when the filter linewidth is much larger than the natural linewidth of
the field, $\Gamma \gg \gamma$. The intensity of the filtered field is
then given by:
\begin{eqnarray}
I_{\omega_i,\Gamma}(t) &=& E_0^2 \frac{\Gamma^2}{2} \mathrm{Re} \int_0^t dt_1 \int_0^{t_1}dt_2\, e^{-\Gamma(t_1+t_2)/2} e^{i [\Delta_i (t_1-t_2) + \phi(t-t_1) -\phi(t-t_2)]} \nonumber \\
&=& E_0^2 \frac{\Gamma^2}{2}  \int_0^t dt_1 \int_0^{t_1}dt_2\, e^{-\Gamma(t_1+t_2)/2} \cos\left[\Delta_i (t_1-t_2) +\Delta\phi(t_1,t_2,t)\right] \nonumber \\
&=& E_0^2 \frac{\Gamma^2}{2}  \int_0^t dt_1 \int_0^{t_1}dt_2\, e^{-\Gamma(t_1+t_2)/2} \left\{ \cos\left[\Delta_i (t_1-t_2)\right]\cos[\Delta\phi(t_1,t_2,t)] \right. \nonumber\\
&&\kern7cm- \left. \sin[\Delta_i(t_1-t_2)]\sin[\Delta\phi(t_1,t_2,t)]\right\}\,,
\label{eq:mean-intensity}
\end{eqnarray}
where $\Delta\phi(t_1,t_2,t) = \phi(t-t_1) -\phi(t-t_2)$. If $\Gamma
\gg \gamma$, the timescale given by the filter linewidth is much
shorter than the natural timescale of the filtered field, and we can
assume $\Delta\phi(t_1,t_2,t) \ll 1$ for those values of $t_1$ and
$t_2$ where the integrand is non-negligible. By expanding to first
order in $\Delta\phi(t_1,t_2,t)$, we obtain:
\begin{equation}
I_{\omega_i,\Gamma}(t)  \stackrel[\Gamma \gg \gamma]{}{\approx}  E_0^2 \frac{\Gamma^2}{2}  \int_0^t dt_1 \int_0^{t_1}dt_2\, e^{-\Gamma(t_1+t_2)/2} \left\{ \cos\left[\Delta_i (t_1-t_2)\right] - \sin[\Delta_i(t_1-t_2)]\Delta\phi(t_1,t_2,t)  \right\} = \langle I_{\omega_i,\Gamma} \rangle + \delta I_{\omega_i,\Gamma}(t)
\end{equation}
where
\begin{equation}
\langle I_{\omega_i,\Gamma} \rangle  = E_0^2 \frac{\Gamma^2}{2}  \int_0^t dt_1 \int_0^{t_1}dt_2\, e^{-\Gamma(t_1+t_2)/2} \cos\left[\Delta_i (t_1-t_2)\right] = E_0^2\frac{(\Gamma/2)^2}{\Delta_i^2 + (\Gamma/2)^2}
\end{equation}
(in agreement with Eq.~\eqref{eq:mean-intensity} in the limit $\Gamma \gg \gamma$) and
\begin{equation}
\delta I_{\omega_i,\Gamma}(t) = - E_0^2 \frac{\Gamma^2}{2}  \int_0^t dt_1 \int_0^{t_1}dt_2\, e^{-\Gamma(t_1+t_2)/2} \sin[\Delta_i(t_1-t_2)]\Delta\phi(t_1,t_2,t)\,.
\end{equation}
Since this equation changes sign when $\Delta_i$ changes sign, we
observe that, in this limit, the fluctuations around the mean value in
opposite sides of the spectrum are perfectly anticorrelated at all
times:
\begin{equation}
  \label{eq:viemar11102705CET2016}
  \delta I_{\omega_0-\omega,\Gamma}(t) = -\delta I_{\omega_0+\omega,\Gamma}(t)\,,
\end{equation}
and $g^{(2)}(\omega_0+\omega,\omega_0-\omega)$ is lower than one:
\begin{equation}
g^{(2)}(\omega_0+\omega,\omega_0-\omega)  = 1 + \frac{\langle \delta I_{\omega_0-\omega}\delta I_{\omega_0+\omega}\rangle}{\langle I_{\omega_0-\omega,\Gamma} \rangle \langle I_{\omega_0+\omega,\Gamma} \rangle} = 1 - \frac{\langle \delta I_{\omega_0+\omega}^2\rangle}{\langle I_{\omega_0+\omega,\Gamma} \rangle^2} < 1\,.
\end{equation}

\subsection*{II.~Frequency correlations of quantum states}

The formal theory of time and frequency resolved correlations is
well-established since the 80s~\cite{cohentannoudji79a, dalibard83a,
  knoll86a, nienhuis93a}.  The two-photon frequency correlations is
expressed as (Eq.~(2) of the text):
\begin{equation}
  \label{eq:marjul29170631CEST2014}
  g_{\Gamma}^{(2)}(\omega_1,t_1;\omega_2,t_2)=\frac{\mean{:\mathcal{T}\big[\prod_{i=1}^2 \hat  E_{\omega_i,\Gamma}(t_i)\hat E^+_{\omega_i,\Gamma}(t_i)\big]:}}{\prod_{i=1}^2\mean{\hat E_{\omega_i,\Gamma}(t_i)\hat E^+_{\omega_i,\Gamma}(t_i)}},
\end{equation}
where
\begin{equation}
\hat E_{\omega_i,\Gamma}(t_i)= \frac{\Gamma}{2}\int_{0}^{\infty}
e^{-i\omega_i t}e^{-\Gamma t/2}\hat E(t_i-t)\,dt
\label{eq:marjul29170601CEST2014}
\end{equation}
is the field of frequency component $\omega_i$ and width $\Gamma$, at
time $T_i$, and $\mathcal{T}$, (resp.~$:$) refer to time
(resp.~normal) ordering. This must be contrasted with the conventional
second-order correlation function:
\begin{equation}
  \label{eq:lundic15191447CET2014}
  g^{(2)}_0=\frac{\sum_{n=0}^\infty n(n-1)\bra{n}\rho\ket{n}}{(\sum_{n = 0}^\infty n \bra{n} \rho\ket{n})^2}\,,
\end{equation}
that only requires the density matrix~$\rho$ to be computed. On the
other hand, for $g^{(2)}(\omega_1,T_1;\omega_2,T_2)$, one needs the
time dynamics even to compute zero-delay coincidences with~$T_1=T_2$
since one has to integrate over time~$t$, as seen in
Eq.~(\ref{eq:marjul29170601CEST2014}). When including the frequency
information, one must therefore specify which dynamics is bringing the
photons from the state towards the detectors that will correlate
them. The fact that such basic physical processes are required to
compute the 2PS shows that it is more fundamental in character than
the conventional $g^{(2)}$. This is similar to early descriptions by
Eberly and W\'odkiewicz~\cite{eberly77a} of photoluminescence spectra
of light beyond the Wiener-Khintchin theorem, that presupposes no
emission. Here too, the necessity to take into account emission and
detection of the photons to define a physical spectrum of light was
pointed out.

The most basic process to bring a photon from the quantum state toward
a detector is spontaneous emission, followed by free propagation
towards the detector that performs the frequency-filtering and
correlation. This provides us with the simplest dynamics to which one
can subject the time evolution of the quantum state of an harmonic
oscillator, as ruled by the master equation for its density matrix:
\begin{equation}
  \label{eq:juesep11094355CEST2014}
  \frac{\partial \rho}{\partial t}=\left[\frac{\gamma_a}2\mathcal{L}_{a}+\frac{\gamma_\phi}2\mathcal{L}_{\ud{a}a}\right](\rho)\,.
\end{equation}
We have also included pure dephasing, for reasons explained in the
main text.  The equation can be integrated in closed form for
$\rho_{n,m}=\langle n|\rho|m\rangle$ which takes the form:
\begin{equation}
  \label{eq:lunnov17184436CET2014}
  \dot{\rho}_{n,m}=-\frac{1}{2}\left[\gamma_a(n+m)+\gamma_\phi(n-m)^2\right]\rho_{n,m}+\gamma_a\sqrt{(m+1)(n+1)}\rho_{n+1,m+1}\,,
\end{equation}
and that can be solved by recurrence, yielding:
\begin{equation}
  \label{eq:juedic11153016CET2014}
  \rho_{n,m}(t) = \sum_{k=0}^{\infty} \rho_{k,m-n+k}(0) \sqrt{\binom{k}{n}\binom{m-n+k}{m}}\left (e^{\gamma_a t}-1\right)^{k-n}  e^{-\left [\gamma_a (2k+m-n)+\gamma_{\phi}(n-m)^2 \right]t/2}\,.
\end{equation}
From $\rho(t)$, one can compute all single-time observables,
such as the population:
\begin{equation}
  \label{eq:juesep11184328CEST2014}
  n(t)=\langle\ud{a}a\rangle(t)=n(0)\exp(-\gamma_a t)\,,
\end{equation}
i.e., simple exponential decay, as expected on physical grounds and
despite the complicated form of the general solution. The two-photon
correlation:
\begin{equation}
  \label{eq:juesep11184455CEST2014}
  g^{(2)}(t)=\frac{\langle\ud{a}\ud{a}aa\rangle(t)}{\langle\ud{a}a\rangle(t)^2}\,
\end{equation}
provides an even simpler and stronger result:
\begin{equation}
  \label{eq:miedic17184244CET2014}
  g^{(2)}(t)=g^{(2)}(0)\,.
\end{equation}
The photon-statistics is constant with time. One can also compute the
two-times correlation function (Eq.~(1) of the main text) through the
quantum regression theorem (demonstration not given), and find a
similarly constrained result:
\begin{equation}
  \label{eq:juesep11184636CEST2014}
  g^{(2)}(t,\tau)=g^{(2)}(0,0)\,.
\end{equation}
This implies, for instance,
$\lim_{\tau\rightarrow\infty}g^{(2)}(t,\tau)\neq1$ for most of the
cases, i.e., photons are always correlated. This is reasonable since
any two photons emitted by the system come from the same and only
initial state which is let to evolve at precisely $t=0$. 

Since we are dealing with many closely related variations of
$g^{(2)}$, we will use the following definition for what is the
central quantity of this work, the zero-delay (coincidence) second
order correlation function:
\begin{equation}
  \label{eq:miedic17184401CET2014}
  g^{(2)}_0\equiv g^{(2)}(t=0,\tau=0)\,.
\end{equation}
This quantity is usually found in the literature written as
$g^{(2)}(0)$.

We now compute the two-photon correlations from an initial state when
including the frequency degree of freedom. To keep the discussion as
fundamental and simple as possible, we consider here the
time-integrated case that disposes of time altogether:
\begin{equation}
  \label{eq:vienov7182502CET2014}
  g^{(2)}_\Gamma(\omega_1,\omega_2)=\frac{\iint_0^{\infty} \mean{n_1(t_1)n_2(t_2)} dt_1dt_2}{\int_0^{\infty} \mean{n_1(t_1)} dt_1\int_0^{\infty} \mean{n_2(t_2)} dt_2}
\end{equation}
where~$n_i=\ud{A_{\omega_i,\Gamma}}A_{\omega_i,\Gamma}$ for~$i=1,2$
are sensors detecting in the corresponding spectral
windows~\cite{delvalle12a}. This is equivalent to letting the
detectors gather statistical information from photons detected at any
time, hence reconstructing the frequency of the photons with full
precision. This quantity is the closest one to what an actual
experiment would perform, although other configurations are possible
(they would bring us to an essentially identical discussion and
conclusions).  Applying Eq.~(\ref{eq:vienov7182502CET2014}) to the
case of free propagation only ($\gamma_a=0$ and $\gamma_\phi=0$) is
pathological because the energy is then exactly determined and
frequency correlations become trivial or ill-defined in terms of
$\delta$ functions. The simplest physically sound dynamics is that of
a free field that, at least, decays ($\gamma_a\neq 0$ and
$\gamma_\phi=0$). One can then compute a physical
frequency-correlation spectrum, and interpret the photons annihilated
by the decay as those detected by the apparatus to register the
information required to compute the correlations. In this case,
corresponding to spontaneous emission of the state, the result is the
same for frequencies as Eq.~(\ref{eq:juesep11184636CEST2014}):
\begin{equation}
  \label{eq:mienov12162459CET2014}
  g^{(2)}_\Gamma(\omega_1,\omega_2)=g_0^{(2)}\,,
\end{equation}
as will be shown in next Section. Including dephasing on top of
the radiative decay ($\gamma_a\neq 0$ and $\gamma_\phi\neq 0$) brings
us to the result:
\begin{equation}
  \label{eq:vienov7190256CET2014}
  g^{(2)}_\Gamma(\omega_1,\omega_2)=g_0^{(2)}\mathcal{F}_{\Gamma}(\omega_1,\omega_2)\,,
\end{equation}
with $\mathcal{F}_{\Gamma}(\omega_1,\omega_2)$ a \emph{boson form
  factor}, which is independent of the quantum state~$\rho$ in which
the system is prepared, and depends only on the dynamics of emission
and detection:
%
%
\begin{multline}
  \label{eq:formfactor}
  \mathcal{F}_{\Gamma}^{(2)}(\omega_1,\omega_2)=
  \Re\Big\{\frac{(\gamma^2+4\omega_1^2)(\gamma^2+4\omega_2^2)}{2\gamma^2(\gamma+2 i \omega_2)} \Big[
  \frac{\gamma+2\gamma_a}{(\gamma+2\gamma_a)^2+4\omega_1^2}+\frac{\gamma_a}{\gamma+2\gamma_a+2i\omega_2}\\
  \times \Big(
  \frac{\gamma+2\gamma_a-i(\omega_1-\omega_2)}{(\gamma+2\gamma_a-2i\omega_1)(\Gamma+\gamma_a-i(\omega_1-\omega_2))}+
  \frac{\gamma+2\gamma_a+i(\omega_1+\omega_2)}{(\gamma+2\gamma_a+2i\omega_1)(2\gamma-\Gamma-\gamma_a+i(\omega_1+\omega_2))}\Big)\Big]\Big\}\\
  +[1\leftrightarrow  2]\,.
\end{multline}
%
with $\gamma=\Gamma+\gamma_a+\gamma_\phi$. The previous result,
Eq.~(\ref{eq:mienov12162459CET2014}), follows from
$\mathcal{F}_{\Gamma}(\omega_1,\omega_2)$ being unity
when~$\gamma_\phi=0$.   The physical reason behind the need of a dephasing mechanism to obtain a non-trivial form factor is explained in the main text.

In general, when the physics goes beyond that of the mere emission
from a quantum state~$\rho$, and involves virtual processes, dressing
of the states, collective emission, stimulated emission and other
types of likewise quantum correlations, the standard Glauber's
correlation $g^{(2)}_0$ does not simply factorize from
$g^{(2)}_\Gamma(\omega_1,\omega_2)$. In such cases, the 2PS offers a
complex landscape of correlations with strong and characteristic
features~\cite{gonzaleztudela13a}, that can be taken advantage of for
distillation~\cite{delvalle13a}, strongly-correlated
emission~\cite{sanchezmunoz14a} and quantum information
processing~\cite{sanchezmunoz14b}. This is however experimentally
outside the scope of this paper which makes the first step in these
exciting directions by confirming Eq.~(\ref{eq:vienov7190256CET2014}).

\subsection*{III.~Calculation of the Boson form factor.}

We now detail the calculation of
$\mathcal{F}_{\Gamma}(\omega_1,\omega_2)$.  Let us assume a quantum
system described by a set of operators~$a$, $\sigma$, etc., acting in
a Hilbert space~$\mathcal{H}$. In second quantization, these operators
define annihilation operators in the Heisenberg picture. All
single-time quantities can be obtained from correlators of the type
$\mean{a^{\dagger\mu}a^{\nu}\sigma^{\dagger\eta}\sigma^{\theta}\ldots}$
with~$\mu$, $\nu$, $\eta$, $\theta$, etc., integers. Let us
call~$\mathcal{O}$ the set of operators the averages of which
correspond to the correlators required to describe the system, i.e.,
$\mathcal{O}$ includes all the sought observables as well as operators
which couple to them through the equations of motion. In the
following, we assume, without loss of generality, that $a$ is the mode
of interest, the correlations of which are to be computed in time and
frequency during its decaying time dynamics.

We define the time-dependent vector $\mathbf{v}$ as:
\begin{equation}
  \label{eq:ThuMar22124609CET2012}
  \mathbf{v} = \left( \begin{array}{c}
      1\\
      \mean{a}(t) \\
      \mean{\ud{a}}(t) \\
      \mean{\ud{a}a}(t) \\
      \vdots
    \end{array}
  \right)\,,
\end{equation}
composed of the mean values of all relevant operators in~$\mathcal{O}$
taken in some order, which will be kept for the remainder of the text
as starting with the sequence $\mathcal{O}=\{1, a, \ud{a},
\ud{a}a,\dots\}$. From the master equation, one can define for
$\mathcal{O}$ a matrix~$M$ which rules the dynamical evolution
of~$\mathbf{v}$: $ \partial_t \mathbf{v}= M \mathbf{v}$, with solution
$\mathbf{v}=e^{Mt}\mathbf{v}_0$ and integrated value $\overline{
  \mathbf{v}}=\int_0^\infty e^{-\lambda t} \mathbf{v} dt =
\int_0^\infty e^{-\lambda t} e^{M t} \mathbf{v}_0 dt =
\frac{-1}{M-\lambda \mathbf{1} } \mathbf{v}_0$. Note that $M$ is
typically singular, i.e., $|M|=0$, therefore we include an exponential
decay~$e^{-\lambda t}$ that forces the dynamics to die away so that we
can enforce $\lim_{t\rightarrow \infty}e^{-\lambda t}\mathbf{v}
=0$. At the end of the calculations we take the limit $\lambda
\rightarrow 0$.

We now consider two sensors~$\varsigma_i$, $i=1,2$ with
linewidths~$\Gamma_i$ coupled to the system with
strength~$\varepsilon_i$ such that the dynamics of the system is
probed but is otherwise left unperturbed. We then introduce a sensing
vector~$\mathbf{w}$ of steady state correlators, by multiplying
$\varsigma_1^{\dagger\mu_1}\varsigma_1^{\nu_1}\varsigma_2^{\dagger\mu_2}\varsigma_2^{\nu_2}$
with the operators in~$\mathcal{O}$:
\begin{equation}
  \label{eq:WedMar21200056CET2012}
  \mathbf{w}[\mu_1\nu_1,\mu_2\nu_2] = \left( \begin{array}{c}
      \mean{\varsigma_1^{\dagger\mu_1}\varsigma_1^{\nu_1}\varsigma_2^{\dagger\mu_2}\varsigma_2^{\nu_2}}(t) \\
      \mean{\varsigma_1^{\dagger\mu_1}\varsigma_1^{\nu_1}\varsigma_2^{\dagger\mu_2}\varsigma_2^{\nu_2}a}(t) \\
      \mean{\varsigma_1^{\dagger\mu_1}\varsigma_1^{\nu_1}\varsigma_2^{\dagger\mu_2}\varsigma_2^{\nu_2}\ud{a}}(t) \\
      \mean{\varsigma_1^{\dagger\mu_1}\varsigma_1^{\nu_1}\varsigma_2^{\dagger\mu_2}\varsigma_2^{\nu_2}\ud{a}a}(t) \\
      \vdots
    \end{array}
  \right)\,,
\end{equation}
where the indices $\mu_i$ and $\nu_i$ take the values 0 or 1. The
integrated quantity is denoted
$\overline{\mathbf{w}}[\mu_1\nu_1,\mu_2\nu_2] =\int_0^\infty
e^{-\lambda t} \mathbf{w}[\mu_1\nu_1,\mu_2\nu_2] dt$. Furthermore, we
also introduce the two-time correlator vector as:
\begin{equation}
  \label{eq:MonSep8160107CEST2014}
  \mathbf{w}'[\mu_1\nu_1,\mu_2\nu_2] = \left( \begin{array}{c}
      \mean{(\varsigma_1^{\dagger\mu_1}\varsigma_1^{\nu_1})(t)(\varsigma_2^{\dagger\mu_2}\varsigma_2^{\nu_2})(t+\tau)} \\
      \mean{(\varsigma_1^{\dagger\mu_1}\varsigma_1^{\nu_1})(t)(\varsigma_2^{\dagger\mu_2}\varsigma_2^{\nu_2}a)(t+\tau)} \\
      \mean{(\varsigma_1^{\dagger\mu_1}\varsigma_1^{\nu_1})(t)(\varsigma_2^{\dagger\mu_2}\varsigma_2^{\nu_2}\ud{a})(t+\tau)} \\
      \mean{(\varsigma_1^{\dagger\mu_1}\varsigma_1^{\nu_1})(t)(\varsigma_2^{\dagger\mu_2}\varsigma_2^{\nu_2}\ud{a}a)(t+\tau)} \\
      \vdots
    \end{array}
  \right)\,,
\end{equation}
with a two-time integral as $\overline{\overline
  {\mathbf{w}}}[\mu_1\nu_1,\mu_2\nu_2]=\iint_0^\infty e^{-\lambda t}
e^{-\lambda \tau} \mathbf{w}'[\mu_1\nu_1,\mu_2\nu_2] dt d\tau$.
Finally, we define two matrices, $T_\pm$, which, when acting on
$\mathbf{v}$ or $\mathbf{w}$, introduce an extra $\ud{a}$ for $T_+$
and an $a$ for $T_-$, keeping normal ordering.  These matrices always
exist, in infinite or in truncated Hilbert spaces (where, if
truncation is to order~$n$, $a^n$ is an operator in~$\mathcal{O}$ but
$a^{n+1}=0$).

In the regime under consideration, the population
$\langle\ud{\varsigma_i}\varsigma_i\rangle\ll1$ and the equations of
motion are valid to leading order in~$\varepsilon_{1,2}$:
\begin{multline}
  \label{eq:FriMar30165907CEST2012}
  \partial_t \mathbf{w}[\mu_1\nu_1,\mu_2\nu_2]= \large\{ M+[(\mu_1-\nu_1)i\omega_1-(\mu_1+\nu_1){\frac{\Gamma_1}2}
  +(\mu_2-\nu_2)i\omega_2-(\mu_2+\nu_2)\frac{\Gamma_2}2]\mathbf{1} \large\}
  \mathbf{w}[\mu_1\nu_1,\mu_2\nu_2]\\
  +\mu_1(i\varepsilon_1T_+)\mathbf{w}[0\nu_1,\mu_2\nu_2]+\nu_1(-i\varepsilon_1T_-)\mathbf{w}[\mu_10,\mu_2\nu_2]
  +\mu_2(i\varepsilon_2T_+)\mathbf{w}[\mu_1\nu_1,0\nu_2]+\nu_2(-i\varepsilon_2T_-)\mathbf{w}[\mu_1\nu_1,\mu_20]\,.
\end{multline}
Since the sensors are empty  at $t=0$ and at $t=\infty(\bar{w}(0
(\infty))\equiv 0$, we easily obtain the solution for
$\overline{\mathbf{w}}$ by formal integration of these equations,
noting that $\int_0^{\infty}e^{-\lambda t} \partial_t \mathbf{w}\,
dt=\lambda \overline{\mathbf{w}}$:
\begin{multline}
  \label{eq:FriMar30202508CEST2012}
  \overline{\mathbf{w}}[\mu_1\nu_1,\mu_2\nu_2]=\frac{-1}{M+[(\mu_1-\nu_1)i\omega_1-(\mu_1+\nu_1)\frac{\Gamma_1}2+(\mu_2-\nu_2)i\omega_2-(\mu_2+\nu_2)\frac{\Gamma_2}2-\lambda]\mathbf{1}}\\
  \times \Big\{
  \mu_1(i\varepsilon_1T_+) \overline{\mathbf{w}}[0\nu_1,\mu_2\nu_2]+\nu_1(-i\varepsilon_1T_-) \overline{\mathbf{w}}[\mu_10,\mu_2\nu_2]
  +\mu_2(i\varepsilon_2T_+) \overline{\mathbf{w}}[\mu_1\nu_1,0\nu_2]+\nu_2(-i\varepsilon_2T_-) \overline{\mathbf{w}}[\mu_1\nu_1,\mu_20]\Big\}
  \,.
\end{multline}
This produces exactly the same type or recursive relation between the
integrated correlators that one has in the case of a steady state
(under continuous pumping)~\cite{delvalle12a}. The only difference
appears when reaching the last vector in the recursive chain, that now
is given by $\overline{\mathbf{w}}[00,00]=\overline{\mathbf{v}}$
instead of the steady state value.

For the two-time correlator, we have:
\begin{multline}
  \label{eq:MonSep8162549CEST2014}
  \partial_{\tau} \mathbf{w}'[\mu_1\nu_1,\mu_2\nu_2]= \large\{ M+[(\mu_2-\nu_2)i\omega_2-(\mu_2+\nu_2)\frac{\Gamma_2}2]\mathbf{1} \large\}
  \mathbf{w}'[\mu_1\nu_1,\mu_2\nu_2]\\
  +\mu_2(i\varepsilon_2T_+)\mathbf{w}'[\mu_1\nu_1,0\nu_2]+\nu_2(-i\varepsilon_2T_-)\mathbf{w}'[\mu_1\nu_1,\mu_20]\,.
\end{multline}
The formal double integration of these equations, noting that
$\iint_0^{\infty}e^{-\lambda t} e^{-\lambda \tau} \partial_{\tau}
\mathbf{w}'\, dt\,d\tau=-\overline{\mathbf{w}} +\lambda
\overline{\overline{\mathbf{w}'}}$, leads to:
\begin{multline}
  \label{eq:MonSep8163931CEST2014}
  \overline{
    \overline{\mathbf{w}'}}[\mu_1\nu_1,\mu_2\nu_2]=\frac{-1}{M+[(\mu_2-\nu_2)i\omega_2-(\mu_2+\nu_2)\frac{\Gamma_2}2-\lambda]\mathbf{1}}\\
  \times \Big\{\mu_2(i\varepsilon_2T_+)
  \overline{\overline{\mathbf{w'}}}[\mu_1\nu_1,0\nu_2]+\nu_2(-i\varepsilon_2T_-)
  \overline{\overline{\mathbf{w'}}}[\mu_1\nu_1,\mu_20]+\overline{\mathbf{w}}[\mu_1\nu_1,\mu_2\nu_2]\Big\}
  \,.
\end{multline}
The final correlator in the recursive chain is given by
$\mathbf{w}'[\mu_1\nu_1,00]=e^{M\tau}\mathbf{w}[\mu_1\nu_1,00]$ and
the integral, therefore, by:
\begin{equation}
  \label{eq:MonSep8181104CEST2014}
  \overline{\overline{\mathbf{w}'}}[\mu_1\nu_1,00]=\frac{-1}{M-\lambda
    \mathbf{1}} \overline{\mathbf{w}}[\mu_1\nu_1,00]\,.
\end{equation}


At this stage we are ready to obtain the single-photon and two-photon
spectra.  The Eberly or time-dependent spectrum of emission of $a$ is
given by the average population of any one of the two sensors, say,
$\mean{n_1}=\mean{\varsigma^\dagger_1\varsigma_1}(t)$. Its equation of
motion reads $\partial_t \mean{n_1} =-\Gamma_1\mean{n_1}+2\Re
(i\varepsilon_1 \mean{\varsigma_1 \ud{a}}(t))$, and with the above
notations, the total integrated spectrum is therefore given by:
\begin{equation}
  \label{eq:WedMar21032339CET2012}
  \overline{\mean{n_1}}=\lim_{\lambda\rightarrow 0}\frac{2}{\Gamma_1}\Re\Big[i\varepsilon_1T_+ \overline{\mathbf{w}}[01,00]\Big]_1\,.
\end{equation}
The subindex in $[\cdot]_1$ means taking the first element of the
resulting vector. Using the solution
Eq.~(\ref{eq:FriMar30202508CEST2012}), the correlator of interest for
the spectrum reads:
\begin{equation}
  \label{eq:WedMar21033037CET2012}
  \overline{\mathbf{w}}[01,0,0]=\frac{-1}{M+[-i\omega_1-\frac{\Gamma_1}{2}-\lambda]\mathbf{1}}(-i\varepsilon_1T_-)\overline{\mathbf{v}}\,.
\end{equation}


The two-photon spectrum follows similarly. The integral of the
intensity correlations between two sensors,
$\mean{n_1(t_1)n_2(t_2)}=\mean{(\varsigma^\dagger_1\varsigma_1)(t_1)(\varsigma^\dagger_2\varsigma_2)(t_2)}$,
is given by $\iint_0^{\infty} \mean{n_1(t_1)n_2(t_2)} dt_1 dt_2
=\lim_{\lambda\rightarrow 0}\overline{\overline{\mean{n_1
      n_2}}}+[1\leftrightarrow 2]$ where
$\overline{\overline{\mean{n_1 n_2}}}=\iint_0^{\infty} e^{-\lambda
  t}e^{-\lambda \tau}\mean{n_1(t)n_2(t+\tau)} dt d\tau$ and
$[1\leftrightarrow 2]$ means to exchange sensor parameters in the
previous expression. The correlator of interest,
$\mean{n_1(t)n_2(t+\tau)}$, with equation of motion:
\begin{equation}
  \label{eq:ThuMar29192644CEST2012}
  \partial_\tau \mean{n_1(t)n_2(t+\tau)}
  =-\Gamma_2\mean{n_1(t)n_2(t+\tau)}+2\Re \Big[ i\varepsilon_2
  \mean{n_1(t)(\varsigma_2 \ud{a})(t+\tau)} \Big]\,,
\end{equation}
relies on the vectors $\mathbf{w}'[11,\mu_2\nu_2]$. In particular,
$\mean{n_1(t)(\varsigma_2 \ud{a})(t+\tau)}$ is the first element of
the vector $T_+\mathbf{w}'[11,01]$. The integrated correlator reads:
\begin{equation}
  \overline{\overline{\mean{n_1
        n_2}}}=\frac{1}{\Gamma_2+\lambda}\Big\{ \overline{\mean{n_1
        n_2}}+2\Re\Big[i\varepsilon_2 T_+ \overline{\overline{\mathbf{w}'}}[11,01]\Big]_1\Big\}
\end{equation}
The one-time integrated correlator is given by:
\begin{equation}
  \label{eq:WedMar21185114CET2012}
  \overline{\mean{n_1n_2}}=\frac{1}{\Gamma_1+\Gamma_2+\lambda}2\Re \Big[ i\varepsilon_2T_+ \overline{\mathbf{w}}[11,01]\Big]_1+[1\leftrightarrow 2]\,,
\end{equation}
while the solution for $\overline{\overline{\mathbf{w}'}}[11,01]$ is:
\begin{equation}
  \overline{\overline{\mathbf{w}'}}[11,01]= \frac{-1}{M+(-i\omega_2-\frac{\Gamma_2}2-\lambda)\mathbf{1}}\Big\{ -i\varepsilon_2 T_- \overline{\overline{\mathbf{w}'}}[11,00]+\overline{ \mathbf{w}}[11,01] \Big\}\,.
\end{equation}
%
 Altogether, using Eq.~(\ref{eq:MonSep8181104CEST2014}), we get the
final expression for the integrated correlations:
  \begin{multline}
    \label{eq:MonSep8182922CEST2014}
    \iint_0^{\infty} \mean{n_1(t_1)n_2(t_2)} dt_1
    dt_2=\lim_{\lambda\rightarrow 0} \frac{2}{\Gamma_2}\Re\Big[
    (i\varepsilon_2 T_+)\Big\{
    \Big[\frac{1}{\Gamma_1}\mathbf{1}+\frac{-1}{M+(-i\omega_2-\frac{\Gamma_2}2)\mathbf{1}}\Big]\overline{
      \mathbf{w}}[11,01]\\
    +\frac{-1}{M+(-i\omega_2-\frac{\Gamma_2}2)\mathbf{1}}
    (-i\varepsilon_2 T_-)\frac{-1}{M-\lambda
      \mathbf{1}}\overline{\mathbf{w}}[11,00]\Big\} \Big]_1+[1\leftrightarrow 2]\,.
  \end{multline}
The required vectors are given by the solutions:
\begin{equation}
  \overline{\mathbf{w}}[11,01]= \frac{-1}{M+(-i\omega_2-\Gamma_1-\frac{\Gamma_2}2-\lambda)\mathbf{1}}  \Big\{ -i\varepsilon_2 T_- \overline{\mathbf{w}}[11,00]-i\varepsilon_1 T_-
  \overline{\mathbf{w}}[10,01]+i\varepsilon_1 T_+ \overline{\mathbf{w}}[01,01] \Big\}\,,
\end{equation}
with:
\begin{subequations}
  \label{eq:ThuMar29195227CEST2012}
  \begin{align}
    &\overline{\mathbf{w}}[11,00]=
    \frac{-1}{M+(-\Gamma_1-\lambda)\mathbf{1}} \Big\{i\varepsilon_1
    T_+ \overline{\mathbf{w}}[01,00]-i\varepsilon_1 T_-
    \overline{ \mathbf{w}}[10,00] \Big\}\,,\\
    &\overline{\mathbf{w}}[10,01]=
    \frac{-1}{M+(i\omega_1-i\omega_2-\frac{\Gamma_1+\Gamma_2}2-\lambda)\mathbf{1}}\Big\{-i\varepsilon_2
    T_- \overline{\mathbf{w}}[10,00]+i\varepsilon_1 T_+
    \overline{\mathbf{w}}[00,01] \Big\}\,,\\
    &\mathbf{w}[01,01]=
    \frac{-1}{M+(-i\omega_1-i\omega_2-\frac{\Gamma_1+\Gamma_2}2-\lambda)\mathbf{1}}\Big\{-i\varepsilon_1
    T_- \mathbf{w}[00,01]-i\varepsilon_2 T_- \mathbf{w}[01,00]
    \Big\}\,.
 \end{align}
\end{subequations}
Finally:
\begin{subequations}
  \label{eq:MonJul16004354CEST2012}
  \begin{align}
    &\overline{\mathbf{w}}[10,00]=
    \frac{-1}{M+(i\omega_1-\frac{\Gamma_1}2-\lambda)\mathbf{1}} i\varepsilon_1 T_+ \overline{\mathbf{v}}\,,\\
    &\overline{\mathbf{w}}[00,01]=\frac{-1}{M+(-i\omega_2-\frac{\Gamma_2}2-\lambda)\mathbf{1}} (-i\varepsilon_2
    T_-) \overline{\mathbf{v}}\,,\\
    &\overline{\mathbf{w}}[01,00]=\frac{-1}{M+(-i\omega_1-\frac{\Gamma_1}2-\lambda)\mathbf{1}} (-i\varepsilon_1
    T_-) \overline{\mathbf{v}}\,.
  \end{align}
\end{subequations}
With this, we get the final formulas to include in
Eq.~(\ref{eq:MonSep8182922CEST2014}):
%
\begin{multline}
  \label{eq:MonSep8191139CEST2014}
 \overline{\mathbf{w}}[11,01] =-i \varepsilon_1^2\varepsilon_2\frac{1}{M+(-i\omega_2-\Gamma_1-\frac{\Gamma_2}2-\lambda)\mathbf{1}}\\
  \times \Big\{ T_- \frac{1}{M+(-\Gamma_1-\lambda)\mathbf{1}}
  \Big( T_+ \frac{1}{M+(-i\omega_1-\frac{\Gamma_1}2-\lambda)\mathbf{1}}T_- + T_-
  \frac{1}{M+(i\omega_1-\frac{\Gamma_1}2-\lambda)\mathbf{1}}  T_+  \Big)\\
  + T_-
  \frac{1}{M+(i\omega_1-i\omega_2-\frac{\Gamma_1+\Gamma_2}2-\lambda)\mathbf{1}}
  \Big( T_- \frac{1}{M+(i\omega_1-\frac{\Gamma_1}2-\lambda)\mathbf{1}}  T_+ + T_+ \frac{1}{M+(-i\omega_2-\frac{\Gamma_2}2-\lambda)\mathbf{1}}T_-  \Big)\\
  +T_+\frac{1}{M+(-i\omega_1-i\omega_2-\frac{\Gamma_1+\Gamma_2}2-\lambda)\mathbf{1}}
  T_-
  \Big(\frac{1}{M+(-i\omega_2-\frac{\Gamma_2}2-\lambda)\mathbf{1}} +\frac{1}{M+(-i\omega_1-\frac{\Gamma_1}2-\lambda)\mathbf{1}} \Big)T_- \Big\}\frac{1}{M-\lambda\mathbf{1}}\mathbf{v}_0\,,
\end{multline}
and 
\begin{multline}
  \label{eq:MonSep8190820CEST2014}
 \overline{\mathbf{w}}[11,00] =-\varepsilon_1^2
 \frac{1}{M+(-\Gamma_1-\lambda)\mathbf{1}}
  \Big( T_+ \frac{1}{M+(-i\omega_1-\frac{\Gamma_1}2-\lambda)\mathbf{1}}T_- + T_-
  \frac{1}{M+(i\omega_1-\frac{\Gamma_1}2-\lambda)\mathbf{1}}  T_+
  \Big) \frac{1}{M-\lambda\mathbf{1}}\mathbf{v}_0\,.
\end{multline}

All the previous derivation has been kept at a general level, that
could be applied to the spontaneous emission of any system. We now
apply it to the case of interest for our previous discussion, namely,
the case of an harmonic oscillator with decay and pure dephasing,
cf.~Eq.~(\ref{eq:juesep11094355CEST2014}).  In this simple case, the
vector $\mathbf{v}$ needed to compute correlators up to second order,
truncates naturally at $\mean{\ud{a}\ud{a}a a}$ with only 9
elements. The corresponding matrix $M$ is diagonal. The time
integrated spectrum of emission reduces to simply:
\begin{equation}
  \label{eq:TueSep9185604CEST2014}
\int_0^{\infty} \mean{n_1(t_1)} dt_1=\varepsilon^2 \frac{2}{\Gamma \gamma_a} \frac{\gamma/2}{(\gamma/2)^2+\omega_1^2}n_0\,,
\end{equation}
with $n_0$ the initial population of the harmonic mode and
$\gamma=\Gamma+\gamma_a+\gamma_\phi$ (we also took the coupling to
sensors and their decay rates equal for simplicity) and taking the
limit $\lambda\rightarrow 0$. The integrated correlations can be
derived similarly, to provide the more complex expression:
\begin{multline}
  \label{eq:TueSep9191350CEST2014}
  \iint_0^{\infty} \mean{n_1(t_1)n_2(t_2)} dt_1
  dt_2=n_0^2g_0^{(2)}\varepsilon^4 \Re\Big\{\frac{8
    }{\Gamma^2\gamma_a^2(\gamma+2 i \omega_2)} \Big[
  \frac{\gamma+2\gamma_a}{(\gamma+2\gamma_a)^2+4\omega_1^2}+\frac{\gamma_a}{\gamma+2\gamma_a+2i\omega_2}\\
  \times \Big(
  \frac{\gamma+2\gamma_a-i(\omega_1-\omega_2)}{(\gamma+2\gamma_a-2i\omega_1)(\Gamma+\gamma_a-i(\omega_1-\omega_2))}+
  \frac{\gamma+2\gamma_a+i(\omega_1+\omega_2)}{(\gamma+2\gamma_a+2i\omega_1)(2\gamma-\Gamma-\gamma_a+i(\omega_1+\omega_2))}\Big)\Big]\Big\}+[1\leftrightarrow
  2]\,,
\end{multline}
which, according to Eq.~(\ref{eq:vienov7190256CET2014}), finally
provides the analytical expression for the frequency-resolved
two-photon spectrum of spontaneous emission of an arbitrary quantum
state of the harmonic oscillator with pure dephasing:
\begin{equation}
  \label{eq:TueSep9193113CEST2014}
  g^{(2)}_\Gamma(\omega_1,\omega_2)=\frac{\iint_0^{\infty} \mean{n_1(t_1)n_2(t_2)} dt_1
    dt_2}{\int_0^{\infty} \mean{n_1(t_1)} dt_1 \int_0^{\infty}
    \mean{n_2(t_2)} dt_2}=g_0^{(2)} \mathcal{F}_{\Gamma}(\omega_1,\omega_2)\,.
\end{equation}
The exact expression \eqref{eq:formfactor} for $\mathcal{F}_{\Gamma}(\omega_1,\omega_2)$
follows straightforwardly from Eqs.~(\ref{eq:TueSep9185604CEST2014})
and~(\ref{eq:TueSep9191350CEST2014}). This form factor is plotted in Fig.~1c of the main text, and fulfils the following
limits: ${\lim_{\Gamma\rightarrow \infty}
\mathcal{F}_\Gamma(\omega_1,\omega_2)=1}$ (we recover the total
integrated correlations when opening the window to include all the
frequencies) and $\lim_{\gamma_\phi\rightarrow 0}
\mathcal{F}_\Gamma(\omega_1,\omega_2)=1$ (without pure dephasing the
2PS lacks any structure in frequency). More notably
$\lim_{\Gamma\rightarrow 0} \lim_{\gamma_\phi\rightarrow
  \infty}\mathcal{F}_{\Gamma}(\omega_1,\omega_2)=1+\delta_{\omega_1,\omega_2}$,
which recovers indistinguishability bunching, the factor $2!$ for
equal frequencies, and otherwise uncorrelated photons, $1$, for
different different frequencies, is scaling the original correlations.

\subsection*{IV.~Dynamics of an out-of-equilibrium polariton condensate.}

While we have dealt above with the 2PS of spontaneous emission
exactly, it is clear that even in this simple case, the exact
calculation is an awkward process. For the situation of our
experiment, which corresponds instead to a steady state, we recourse
to numerical calculations. Importantly, however, the two situations
are not extremely different from a physical point of view. The
emission indeed corresponds to spontaneous emission from a state whose
coherence depends on the degree of condensation. The final
phenomenology is extremely similar. We describe our system
theoretically by the following minimal model which accounts for all
the key ingredients of the experiments:
\begin{equation}
  \label{eq:lunmay19170025CEST2014}
  \frac{\partial \rho}{\partial t}=\left[\frac{\gamma_a}2\mathcal{L}_{a}+\frac{\gamma_b}2\mathcal{L}_{b}+\frac{P_b}2\mathcal{L}_{\ud{b}}+\frac{P_{ba}}2\mathcal{L}_{\ud{a}b}\right](\rho)\,,
\end{equation}
where $\rho$ is the combined reservoir-condensate density matrix
defined on the Hilbert space of two bosonic fields, since we describe
both the BEC and the exciton reservoir by two harmonic modes $a$ and
$b$, which obey bosonic algebra $[c,\ud{c}]=1$, with $c=a,b$. In the
rotating frame of the frequency of the condensate, the dynamics is
purely dissipative. Both modes lose particles, with decay rate
$\gamma_c$, described by Lindblad terms:
$\sum_{O=a,b}\frac{\gamma_O}2\mathcal{L}_{O} (\rho)$, where
$\mathcal{L}_{O} (\rho)=2O\rho \ud{O}-\ud{O}O\rho-\rho \ud{O}O$.  The
excitation is through the incoherent injection of reservoir excitons
at a rate $P_b$ with the accompanying Lindblad
term~$\frac{P_b}2\mathcal{L}_{\ud{b}}(\rho)$. The transfer of
particles from the reservoir to the condensate, typically assumed to
be phonon mediated, is described by the incoherent relaxation
mechanism from $a$ to $b$, described by a crossed Lindblad term
$\frac{P_{ba}}2\mathcal{L}_{\ud{a}b}(\rho)$~\cite{holland96a}. In an
open system, such a reduced system is enough to capture the physics of
condensation that otherwise requires a macroscopic reservoir with $N$
states and $N\rightarrow\infty$ to achieve coherence
buildup~\cite{laussy12a}. This model has the minimum, but also all,
ingredients to explain the core physical processes that takes place
within our experimental conditions. It accounts successfully for,
e.g., line narrowing and transition to lasing/condensation of the mode
$a$ when the pumping $P_b$ is high enough, to all orders of the
condensate field correlators $N_{ab}[n,0]$, where
$N_{ab}[n,m]=\mean{(\ud{a})^n a^n (\ud{b})^m b^m}$ with
$n,m\in\mathbf{N}$ form a closed set under the dynamics of
Eq.~(\ref{eq:lunmay19170025CEST2014})~\cite{delvalle09c}. It is
therefore also a sound model to compute theoretically the
frequency-resolved correlations.  The zero-time delay dynamics is
easily obtained:
\begin{multline}
  \dot{N}_{ab}[n,m]=-\Big[n \gamma_a+m \big(\gamma_b-P_b+P_{ba}\big)+nm P_{ba}\Big] N_{ab}[n,m]\\
  +n^2 P_{ba}  N_{ab}[n-1,m+1] + n P_{ba} N_{ab}[n,m+1]+P_{b} m^2 N_{ab}[n,m-1] -m P_{ba} N_{ab}[n+1,m]\,.
\end{multline}
Integrating these equations, it is possible to calculate, e.g., the
condensate population, $n_a=N_{ab}[1,0]$, the unnormalized second
order correlation function at zero delay~$G^{(2)}(\tau=0)=N_{ab}[2,0]$
or any other single time correlator.  In particular, the steady state
is obtained by setting $\dot{N}_{ab}[n,m]=0$ and solving the system of
linear equations, which is finite when truncating to a large enough
number of excitations. It is well known, and is straightforwardly
shown, that $g^{(2)}$ goes from values above~1, when $P_b\ll
\gamma_{b,a}$, to $1$ when $P_b\gg \gamma_{b,a}$, corresponding to a
coherence buildup that accompanies condensation with $n_a\gg1$ and
triggering a dynamics of relaxation dominated by stimulated
emission~\cite{laussy12a}.

By following a similar procedure as in the previous section but for
the case of a steady state~\cite{delvalle12a} and the Liouvillian of
Eq.~(\ref{eq:lunmay19170025CEST2014}) we can compute (now numerically)
the 2PS in this case. The result is shown in Fig.~2b of the main text
and is indeed qualitatively very similar to the spontaneous emission
case of a coherent state (where~$g^{(2)}(0)=1$). In fact, the density
plots may appear the same, but one can check by a more careful
analysis that they are not exactly identical. The physics, however,
has the same interpretation: a state with a given $g^{(2)}(0)$
correlations emits photons which, if correlated in frequencies,
exhibit an overall bunching when overlapping in time and frequencies
and antibunching when overlapping in time but distinguished in
frequencies.

\bibliography{Sci,books,arXiv}

\end{document}